\RequirePackage{fix-cm}
\documentclass[smallextended]{svjour3}       

\smartqed  
\usepackage[hyphens]{url}
\usepackage{MnSymbol,wasysym}
\usepackage[]{natbib}
\usepackage{algorithmic}
\usepackage{multirow}
\usepackage{graphicx}
\usepackage{textcomp}
\usepackage[table,xcdraw]{xcolor}
\usepackage{tcolorbox}
\usepackage{booktabs}
\usepackage[hidelinks]{hyperref}
\def\BibTeX{{\rm B\kern-.05em{\sc i\kern-.025em b}\kern-.08em
    T\kern-.1667em\lower.7ex\hbox{E}\kern-.125emX}}

\def\mybar#1{
  {\color{gray}\rule{#1cm}{8pt}}}
  
\graphicspath{{figs/}}

\begin{document}

\begin{sloppy}

\title{GitHub Discussions:\\ An Exploratory Study of Early Adoption}

\author{Hideaki Hata \and
        Nicole Novielli \and
        Sebastian Baltes \and
        Raula Gaikovina Kula \and
        Christoph Treude
}


\institute{Hideaki Hata \at
              Shinshu University \\
              \email{hata@shinshu-u.ac.jp}           
           \and
           Nicole Novielli \at
              University of Bari \\
              \email{nicole.novielli@uniba.it}
           \and
           Sebastian Baltes \at
              QAware GmbH, University of Adelaide \\
              \email{research@sbaltes.com}
           \and
           Raula Gaikovina Kula \at
              Nara Institute of Science and Technology \\
              \email{raula-k@is.naist.jp}
           \and
           Christoph Treude \at
              University of Melbourne \\
              \email{christoph.treude@unimelb.edu.au}
}

\date{Received: date / Accepted: date}

\maketitle

\begin{abstract}
Discussions is a new feature of GitHub for asking questions or discussing topics outside of specific Issues or Pull Requests. Before being available to all projects in December 2020, it had been tested on selected open source software projects. To understand how developers use this novel feature, how they perceive it, and how it impacts the development processes, we conducted a mixed-methods study based on early adopters of GitHub discussions from January until July 2020. We found that: (1) errors, unexpected behavior, and code reviews are prevalent discussion categories; (2) there is a positive relationship between project member involvement and discussion frequency; (3) developers consider GitHub Discussions useful but face the problem of topic duplication between Discussions and Issues; (4) Discussions play a crucial role in advancing the development of projects; and (5) positive sentiment in Discussions is more frequent than in Stack Overflow posts. Our findings are a first step towards data-informed guidance for using GitHub Discussions, opening up avenues for future work on this novel communication channel.
\end{abstract}

\section{Introduction}


As part of the 2020 edition of GitHub's Satellite conference in early May, the platform announced a number of new features, including Codespaces, GitHub Private Instances, and GitHub Discussions. The latter were touted as an alternative to GitHub Issues and Pull Requests, stating that ``software communities don't just write code together. They brainstorm feature ideas, help new users get their bearings, and collaborate on best ways to use the software''.\footnote{\url{https://github.blog/2020-05-06-new-from-satellite-2020-github-codespaces-github-discussions-securing-code-in-private-repositories-and-more/\#discussions}} 
In their announcement, GitHub argued that the linear format of Issues and Pull Requests---while well suited for merging code---is not suitable for creating a community knowledge base. With the goal of filling this gap, GitHub Discussions was released in May 2020 in a closed beta phase for selected public repositories.
In December 2020 during the GitHub Universe conference, the platform announced that GitHub Discussions was released as a public beta feature.\footnote{\url{https://github.blog/2020-12-08-new-from-universe-2020-dark-mode-github-sponsors-for-companies-and-more/\#discussions}}

Originally, this feature was requested by software developers back in 2016.\footnote{\url{https://github.com/dear-github/dear-github/issues/44}}
As such requests show, there are questions about the effectiveness of the communication channels that developers currently have at their disposal, in particular for open-ended tasks such as brainstorming. Discussions is one of the first attempts at addressing this on the GitHub platform.
As previous work reported, innovative services could attract contributors~\citep{10.1109/CHASE.2015.9}.
%
GitHub's announcement drew the attention of the software developer community which questioned the role of the new feature in the existing landscape of question-and-answer (Q\&A) forums, mailing lists, issue trackers, and others. For example, in a popular Hacker News thread\footnote{\url{https://news.ycombinator.com/item?id=22388639}} 
debating the announcement, opinions ranged from ``GitHub is slowly eating the surrounding cities [...] now Discussions aims for a piece of Stack Overflow pie'' to ``This should make Github much more pleasant once it spreads across most projects''. In particular the relationship between the new Discussions feature and the Q\&A forum Stack Overflow garnered interest, with some users expecting a ``reduction'' in Stack Overflow and others arguing that Stack Overflow ``is literally anti-discussion, the whole point is it is about questions and answers, not discussion [...] That isn't to say discussion isn't useful or valuable, it absolutely is, it is just a completely different thing''.

The collaborative and participatory nature of software development is shaped by the communication channels that are used by software developer communities of practice~\citep{storey2016social}. At the same time, these communities will design their own social protocols of how to use tools and adapt and repurpose them over time~\citep{giuffrida2013empirical}. Our work 
follows in a long line of work on how software developers use and appropriate communication channels, ranging from discussions during code review~\citep{TsayFSE14} and on Stack Overflow~\citep{Beyer2020} to mailing lists~\citep{guzzi2013communication}.
Here we aim at understanding whether it is worth adding yet another communication channel. Our focus is not on ``the feature'' but about how it is used for collaborative knowledge-building and how its usage differs from existing channels.

As the software engineering research community is often criticized for being slow in responding to trends, we monitored the closed beta phase to provide data-driven feedback for early adopters who decide to use the feature now that it is publicly available.
In this paper, we report the results of the first comprehensive mixed-methods study on how software developers utilize the GitHub Discussions feature during the closed beta period, what they are using it for and who uses it,
using qualitative and quantitative analysis, online surveys, and sentiment analysis.
We also investigate how these early adopters were initially planning to use it and how they now perceive its usefulness and its place in the complex landscape of existing communication channels and artefacts. We further study the feature's impact on the development process by analysing Discussions' interplay with GitHub Issues and Pull Requests, and we investigate how GitHub Discussions compares to Stack Overflow threads in terms of their sentiment and toxicity.

Based on analysing 7,116 discussions with 29,136 posts from 92 GitHub projects,
we find that software developers use the new feature to discuss errors, discrepancies and reviews, and that most discussions were started by project users. Developers initially envisioned using the new feature for question-answering, idea sharing, and community engagement, and appreciate the integration of discussions into the rest of GitHub while highlighting potential issues related to redundancy. Discussions can drive development by triggering the creation of new Issues and Pull Requests. Compared to Stack Overflow threads, positive sentiment is more frequent in GitHub Discussions.

As GitHub Discussions is now available to all projects hosted on the platform, our findings can guide development teams in their decision on whether and how to adopt the feature by establishing how it is being used by early adopters and reporting the experiences of those using it. Our insights further aim to improve the practices developers follow when contributing to GitHub Discussions and guide GitHub and other platforms on how the feature can be improved, in particular in terms of reducing redundancy. Our work also introduces a novel communication channel to the software engineering literature, opening up venues for future work on GitHub Discussions and their interplay with other channels and artefacts in the complex software development landscape.



The remainder of the paper is organized as follows. In the next section, we introduce the GitHub Discussions feature in detail, before we present our research questions and data collection approach in Sections~\ref{sec:RQs} and~\ref{sec:data}. In Section~\ref{sec:methodology}, we describe the methodology used to address our research questions. Then we present our results in Section~\ref{sec:findings} and discuss limitations in Section~\ref{sec:threats}. Finally, we position our work in the context of related work (Section~\ref{sec:related}) and conclude by envisioning possible challenges for future research based on our findings~(Section~\ref{sec:conclusion}).

\begin{figure*}
  \centering
  \includegraphics[width=\linewidth]{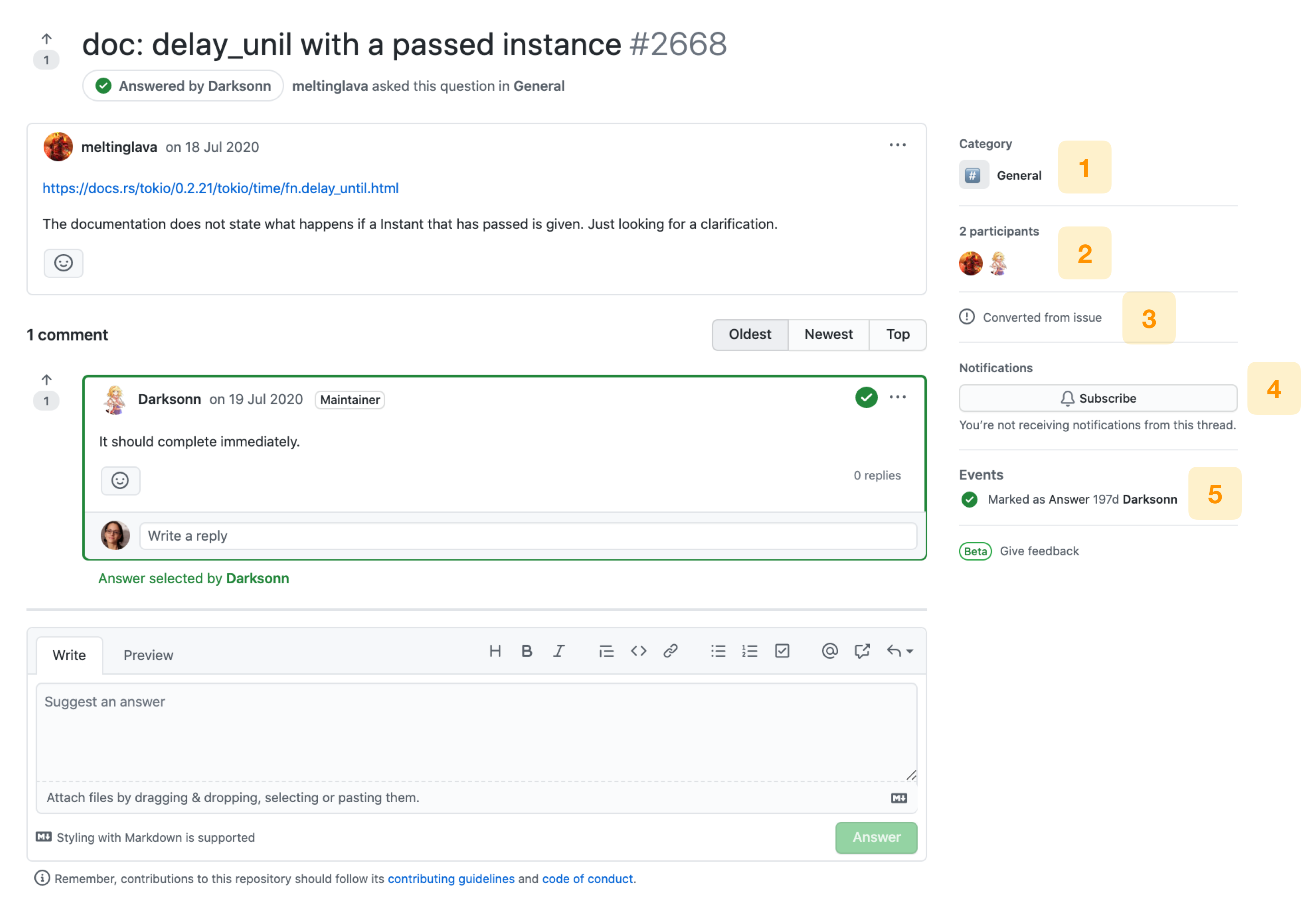}
  \caption{The GitHub Discussions feature: categories, participants, relations, notifications, and events. 
  }
  \label{fig:GHDiscussion}
\end{figure*}

\section{GitHub Discussions}
GitHub Discussions is a new feature for GitHub specifically designed to support communication for the members of an open source project community. The forum is designed to host conversations that, unlike issues, are not directly related to code development and, thus, are not supposed to be included in the project boards. Nevertheless, the official GitHub guidelines\footnote{\url{https://docs.github.com/en/discussions/quickstart}} recommend to clearly define a code of conduct for each project to guide contributors towards effective communication. 

The GitHub Discussion interface is structured in a Question-Answering fashion (see Figure~\ref{fig:GHDiscussion}). However, there are several differences with respect to popular QA-technical forums, such as Stack Overflow, which we highlight in the following. 
Differently from Stack Overflow, a GitHub Discussion can serve different communicative intentions, beyond question-answering. For example a Discussion can be started to share information or make announcements, report errors or discuss about potential evolution of a software project. For example, Figure~\ref{fig:GHDiscussion} depicts a thread starting from a conceptual question aimed at understanding the code behavior.

Discussion are organized according to categories (highlighted as item number 1 in the figure), such as `General', `Q\&A', `Ideas', and so on. Differently from what happens in Stack Overflow tags, where the information seeker can either reuse or create a new tag at the moment of question writing, in GitHub Discussions categories are created by either repository owners and contributors with write access. Participants (2) of discussions can use the most relevant categories among the existing ones in order to group and organize threads. 

Discussion can either starts from scratch or from a previously existing issue. The web interface includes this information (3) together with the number of participants in the discussion. Similarly to other communication in GitHub, one can subscribe to receive notifications (4). 
Similarly to Stack Overflow, it is possible to flag an answer as answering the original question. However, differently from Stack Overflow where the question author is the only one allowed to flag an answer as accepted, in GitHub Discussion this can be done by other people as well, as shown in Figure~\ref{fig:GHDiscussion} (5).

\section{Research Questions}
\label{sec:RQs}
The main goal of the study is to gain insights into the use of GitHub Discussions. 
Based on this goal, we constructed three main research questions to guide our study.
We present these main questions and sub-questions, along with their motivation.

\begin{description}
\item[\textbf{RQ1}] \textit{How have GitHub Discussions been adopted?}
  \begin{description}
  \item[\textbf{RQ1.1}] \textit{What are GitHub Discussions about?}
  \item[\textbf{RQ1.2}] \textit{Who contributes to GitHub Discussions?}
  \item[\textbf{RQ1.3}] \textit{What is the relationship between project characteristics and GitHub Discussion adoption?}
  \end{description}
\end{description}
The motivation of our first main research question (\textbf{RQ1}) is to understand the kind of discussions that occur in GitHub Discussions.
We use a qualitative method to categorize the discussions by the content and the individuals that contribute to the discussions.

\begin{description}
\item[\textbf{RQ2}] \textit{What reasons do developers have to adopt GitHub Discussions?}
  \begin{description}
  \item[\textbf{RQ2.1}] \textit{Why do projects adopt GitHub Discussions?}
  \item[\textbf{RQ2.2}] \textit{How do developers perceive GitHub Discussions?}
  \end{description}
\end{description}
Our second main research question (\textbf{RQ2}) requires an analysis of the different perceptions of why GitHub Discussions is being used and how developers feel about the feature.

\begin{description}
\item[\textbf{RQ3}] \textit{How do GitHub Discussions relate and compare to other communication channels?}
  \begin{description}
  \item[\textbf{RQ3.1}] \textit{What is the impact of GitHub Discussions on other channels?}
  \item[\textbf{RQ3.2}] \textit{How does the sentiment differ compared to Stack Overflow posts?}
  \end{description}
\end{description}
GitHub Discussions does not exist in isolation, but are part of a network of a myriad of other artefacts.
Our third main research question (\textbf{RQ3}) explores the impact of GitHub Discussions on other artefacts and compares emotional styles with an existing communication channel.
The key motivation is that we would like to understand the potential roles of GitHub Discussions for a software project.

\section{Data Collection}
\label{sec:data} 

In this section we describe our data collection methodology, and present our replication package.
When we initiated this research project, the GitHub REST API\footnote{\url{https://docs.github.com/en/rest}} did not support the new Discussions feature yet.
In addition, when we asked the GitHub support for projects that have access to the beta feature, they replied that they could not disclose that information.
Thus, to determine which projects already use the feature and to collect metadata and posts for analysis, we had to rely on a custom web scraper we built (available on GitHub\footnote{\url{https://github.com/sbaltes/github-retriever/}}).
In a first step, we utilized the API to retrieve the most popular projects according to their number of stargazers.
Previous work has shown that this approach has a high precision in retrieving engineered software projects~\citep{DBLP:journals/ese/MunaiahKCN17}.
We assumed that more popular projects have a larger user base and are more likely to adopt Discussions early on.
Our goal was to retrieve at least 10,000 projects.
Due to API limitations, we had to execute multiple queries and then merge the data, yielding the 10,899 most popular GitHub projects as of June 26, 2020.
Next, we used our web scraper to determine which GitHub features, including Discussions, Issues, and Pull Requests, are activated in those projects.
Many projects used Pull Requests (10,896), Actions (10,744), and Issues (10,432).
The Projects feature was less common (8,997), followed by Wikis (4,071) and the new Discussions feature (99).
After manually checking the projects using Discussions, we excluded one not being used for software development.

For the remaining 98 projects, we scraped all discussions, including metadata such as title, state (answered or unanswered), author, timestamp, and whether the discussion thread was converted from an existing Issue.
For individual posts, we stored author, timestamp, 
whether a post was part of the selected answer (posts can be nested in discussions), and the HTML content.

We executed several data collection runs.
The dataset we used to answer \textbf{RQ1} and \textbf{RQ2} was collected on July 22 and contains 7,116 discussions with 29,136 posts from 92 projects; the dataset used to answer \textbf{RQ3} was collected on August 1 and contains 9,234 discussions with 32,714 posts from 92 projects.
The earliest discussion thread in our dataset had started in January 2020.
Note that some projects had the Discussions feature activated, but no discussion threads yet.
Further, some projects converted existing Issues into Discussion threads: In the August 1 dataset, 1,104 discussions (12\%) with 5,885 posts (18\%) were converted from Issues, as opposed to 1,007 (14\%) discussions with 5,401 posts (19\%) in the older one.

\subsection{Online Appendix}

Our appendix containing results of our qualitative analyses is available on GitHub\footnote{\url{https://github.com/sbaltes/GHDiscussions}} and Zenodo~\citep{hideaki_hata_2021_4527166}. The appendix includes files with links to the full Discussions on GitHub. We further published the scripts we used to collect the data not available via the GitHub API.\footnote{\url{https://github.com/sbaltes/github-retriever}}



\section{Methods}
\label{sec:methodology}

This section describes our mixed-methods procedure including a qualitative analysis, a quantitative analysis, a survey, and a semantic analysis.

\subsection{Qualitative Analysis}

To understand what kind of discussions occurred in GitHub Discussions (\textbf{RQ1.1}),
investigate the relationship between project characteristics and GitHub Discussion adoption (\textbf{RQ1.3}),
get an initial view of developer perceptions of the GitHub Discussions feature and to understand why projects adopt GitHub Discussions (\textbf{RQ2.1}),
and investigate how GitHub Discussions affects the projects that they belong to and in particular understand their impact on other channels and artefacts (\textbf{RQ3.1}),
we manually annotate target instances, with multiple coders, in multiple iterations.

\paragraph{Discussion Categories (RQ1.1)}
We conduct a qualitative analysis of statistically representative samples of two groups of discussions, namely, discussions converted from Issues and not from Issues (started originally at GitHub Discussions). We have 1,007 and 6,109 discussions 
for the two groups, respectively. 
We randomly sampled a statistically representative number of discussions from each discussion group, ensuring that our findings regarding percentages within each sample would generalize to the entire groups with a confidence level of 95\% and a confidence interval of 5. We obtained 278 (from Issues) and 361 Discussions (not from Issues) for our samples, for a total of 639 discussions overall.
Our codes for discussion categories were informed by a taxonomy of question categories to understand kinds of questions on Stack Overflow~\citep{Beyer2020}. 
Since we found that their codes did not cover all types of discussions on GitHub Discussions, we developed our coding schema by modifying certain codes. The following list shows the 13 categories along with their descriptions. The top seven categories are based on the existing taxonomy~\citep{Beyer2020}: some are exactly the same and the others are slightly modified. Four categories emerged from our analysis, in addition to \textit{Other} and \textit{404}. These four categories are unique discussion categories in GitHub Discussions compared to questions on Stack Overflow.
\begin{itemize}
    \item \textit{Errors}: This is said to be equivalent to the categories \textsf{Error}~\citep{10.1145/1985793.1985907} and \textsf{Exception handling}~\citep{10.1145/2901739.2901750}. Discussions of this category contain exceptions and stack traces and ask for help in fixing errors or understanding what the exceptions mean.
    \item \textit{Discrepancy}: This contains the categories \textsf{Discrepancy}~\citep{10.1145/1985793.1985907}, \textsf{Do not work}~\citep{10.5555/2487085.2487098}, and \textsf{What is the problem?}~\citep{10.1145/2901739.2901750}. Such discussions contain questions about problems or unexpected behavior where the questioners have no clue how to solve them.
    \item \textit{Review}: This category merged \textsf{Decision help} and \textsf{Review}~\citep{10.1145/1985793.1985907}, \textsf{How/why something works}~\citep{10.5555/2487085.2487098}, \textsf{Better solution}~\citep{10.1145/2901739.2901750}, and \textsf{What}~\citep{10.1007/s10664-015-9379-3}. Discussions of this category contain code snippets and ask for better solutions, help to make decisions, or to review their code snippets.
    \item \textit{Conceptual}: This is equivalent to the category \textsf{Conceptual}~\citep{10.1145/1985793.1985907} and related to the categories \textsf{Why} and \textsf{Is it possible?}~\citep{10.1145/2901739.2901750}, \textsf{What}~\citep{10.1007/s10664-015-9379-3}, and \textsf{How/why something works}~\citep{10.5555/2487085.2487098}. These discussions mainly ask high-level questions, such as the limitations of functionalities, underlying concepts (design patterns, architectural styles, etc.), and background information. 
    \item \textit{Usage}: The taxonomy contains the category \textsf{API usage}~\citep{Beyer2020}, which is related to the categories \textsf{How to implement something} and \textsf{Way of using something}~\citep{10.5555/2487085.2487098} and \textsf{How-to}~\citep{10.1145/2901739.2901750,10.1145/1985793.1985907}.
    Since we observed questions of usages not limited to APIs, we extended this category to general usages. Discussions of this category contain questions asking for concrete instructions of specific functionalities, which is different from questions in \textit{Discrepancy}.
    \item \textit{Learning}: This is related to the category \textsf{Learning a language/technology}~\citep{10.5555/2487085.2487098}.
    In these discussions, the questioners ask for documentation or tutorials to learn the software of the projects. Instead of asking for solutions or instructions on how to do something, they aim at asking for resources of documentation, tutorials, or examples, to learn on their own.
    \item \textit{Versions}: The taxonomy contains the category \textsf{API change}~\citep{Beyer2020}, which is equivalent to the category \textsf{Version}~\citep{10.1145/2901739.2901750}.
    Discussions in this category are not limited to API changes but are related to any questions that arise because of different versions of the software or environmental settings.
    \item \textit{Plans}: Discussions of this category ask for future plans of software releases or development processes.
    \item \textit{Announcement}: These discussions are used for announcements about specific events of the software (e.g., future updates, releases).
    \item \textit{Information}: Discussions in this category provide general information, which is not related to events of the software.
    \item \textit{Recruitment}: These discussions indicate offers of job vacancies or recruiting contributors to the teams.
    \item \textit{Other}: Discussions that do not fit the above categories.
    \item \textit{404}: Discussions which had been removed when we accessed them.
\end{itemize}
Our annotation guidelines do not allow for multiple categories. We annotate the categories based on the initial posts of the threads.
The annotation was conducted by three of the authors of this paper. We conducted an initial pilot annotation on two subsets of 30 discussions from the two groups (from Issues and not from Issues). The three coders read and discussed the annotation taxonomy in order to reach a shared understanding of the codes. The three coders labeled individually all items in the two pilot sets of 30 discussions.
After this first iteration, we computed the inter-rater agreement to assess the reliability of the annotation approach. Specifically, we computed Fleiss' $k$ index, which extends Cohen’s $k$ for evaluating the inter-rater level of agreement between two or more raters, in presence of a categorical variable~\citep{Fleiss}. It expresses the degree to which the observed proportion of agreement among raters exceeds what would be expected if all raters made their ratings completely randomly.
We obtained an observed agreement of 50\% and $k$ = .46, denoting moderate agreement~\citep{Viera2005}. In order to consolidate the annotation guidelines, the three raters discussed and resolved the disagreement cases. We then incorporated the insights derived from the discussion phase into the annotation guidelines and repeated the pilot annotation. After this second round, we observed substantial agreement, with $k$ = .63 and observed agreement = 66\%.
Given the reliability of the annotation guidelines, two of the raters individually completed the annotation for the remaining cases in the two samples.

\paragraph{Project Characteristics (RQ1.3)}
For software domains, we adopt the following five domains presented in a previous study of GitHub repositories~\citep{7816479}. 
While there were six domains in the previous study, since this study focuses on repositories for software development, we excluded ``Documentation,'' which represents repositories with documentation, tutorials, and source code examples. In this analysis, all repositories could be classified into the five domains.
 \begin{itemize}
    \item \textit{Web libraries and frameworks}: Software for web development. 
    \item \textit{Software tools}: Systems that support software development tasks, like IDEs, package managers, and compilers. 
    \item \textit{Application software}: Systems that provide functionalities to end-users, like browsers and text editors. 
    \item \textit{Non-web libraries and frameworks}: Frameworks not intended for web development. 
    \item \textit{System software}: Systems that provide services and infrastructure to other systems, like operating systems, middleware, servers, and databases. 
\end{itemize}
Three authors independently coded 98 projects. 
We conducted a pilot analysis in order to assess the reliability of our annotation schema. The three raters independently labeled 28 projects from our sample, achieving a substantial inter-rater agreement ($k$ = .61, observed agreement = 69\%). Thus, the three raters completed the annotation of the software domains for the remaining projects, obtaining again a substantial agreement ($k$ = .64, observed agreement = 71\%). Overall, we observed a $k$ = .63 and observed agreement = 71\%. We assign the final domain label to all repositories in our sample based on the majority agreement among the three raters. On four projects, a majority agreement was not reached. Thus, the three raters assigned the domain labels to these cases after a discussion.

\paragraph{Initial Discussions (RQ2.1)}
We conduct a qualitative analysis of the first discussion thread of each project in our data set since we had anecdotally observed that the first thread was often used for meta-discussions, i.e., discussions about the Discussion feature and social negotiation of how it should be used~\citep{giuffrida2013empirical}. One author developed a coding schema based on the inspection of 30 randomly selected first threads, and another author annotated the same threads to establish the reliability of the coding schema via inter-rater agreement. We allowed multiple codes per discussion thread. Both annotators achieved perfect agreement in 23/30 cases (77\%) and partial agreement in another 4/30 cases (13\%), and they disagreed in the remaining 3/30 cases (10\%). Based on this satisfactory agreement, one author then annotated the remaining threads. The coding schema we established consisted of the following codes:
\begin{itemize}
    \item \textit{Question-answering}: Several projects indicated that the Discussion feature could be used for asking and answering questions, e.g., ``I'm trying out GitHub Discussions in the hopes that folks will be encouraged to both ask questions and give answers. Remember, there are no stupid questions!''\footnote{\url{https://github.com/BurntSushi/ripgrep/discussions/1552}}
    \item \textit{Idea sharing}: Sharing of ideas was also mentioned by several projects, although few projects provided details of how such sharing could be structured.
    \item \textit{Community engagement}: Projects also mentioned using GitHub Discussions to engage with their broader community, e.g., ``We want to try using GitHub's new discussion feature to structure conversations that may not quite be an Issue, but would benefit from community involvement and searchability.''\footnote{\url{https://github.com/linkerd/linkerd2/discussions/4347}}
    \item \textit{Decluttering issue tracker}: A few projects explicitly mentioned the relationship between GitHub Discussions and GitHub Issues, suggesting that Discussions could help declutter the issue tracker, e.g., ``Unsure whether what you're experiencing is a bug or not? Post it here first! if it is indeed a bug, moving it to Issues once it's confirmed is an easy task.''\footnote{\url{https://github.com/containrrr/watchtower/discussions/567}}
    \item \textit{Information resource building}: Projects also mentioned using Discussions as a way of building an information resource, which is in line with GitHub's original intent behind the Discussions feature.
    \item \textit{Feature requests}: Discussing new features was also mentioned as a use case of GitHub Discussions.
    \item \textit{Thank-you}: In one case, the first discussion thread was used to say thank-you to everyone involved in the project.
    \item \textit{Testing the feature}: Several development teams used their first GitHub Discussion to test the new functionality, in particular threaded replies.
    \item \textit{Not meta}: We used the code ``not meta'' to indicate cases where the first Discussion thread had not been used to discuss the feature itself.
\end{itemize}

\paragraph{Impact on Channels (RQ3.1)}
To study the impact of GitHub Discussions on other channels, we collect all links from GitHub Discussions to Issues and Pull Requests in the same repository. To focus on those artefacts that could actually have been affected by a Discussion, we limit our analysis to Issues and Pull Requests which had been opened or closed after the Discussion started. We found 577 such links in our dataset.
To investigate whether these links are indicators of GitHub Discussions affecting other artefacts, we manually analysed a statistically representative and random sample of 231 links to determine the relationship between the Discussion and the Issue or Pull Request. Two authors annotated 20 links separately (Cohen's $k$ = 0.93), and based on the near perfect agreement, one author annotated the remaining links. We used the following codes:
\begin{itemize}
     \item \textit{Newly open}: An Issue or Pull Request was opened as a result of a GitHub Discussion. For example, Pull Request \#498\footnote{\url{https://github.com/alpinejs/alpine/pull/498}} was opened as a result of Discussion \#489\footnote{\url{https://github.com/alpinejs/alpine/discussions/489}} in the alpinejs/alpine repository. The discussion started with a question around potential library use, and after ten discussion messages, a contributor announced the corresponding Pull Request which has since been merged.
    \item \textit{Deepen discussion}: In some cases, GitHub Discussions was used to deepen the discussions related to an Issue or Pull Request, in many cases exploring the broader implications of these artefacts.
    \item \textit{Reference}: In other cases, the Discussion thread did not directly affect an Issue or Pull Request, but referenced it for traceability purposes. For example, Discussion \#4383\footnote{\url{https://github.com/linkerd/linkerd2/discussions/4383}} in the linkerd/linkerd2 repository provides an update on the project roadmap, referencing five Issues or Pull Requests as part of the update.
\end{itemize}

\subsection{Quantitative Analysis (RQ1.2)}

To identify the types of participants who contribute to GitHub Discussions (\textbf{RQ1.2}), we investigated the roles of all discussion participants in our dataset using the GitHub REST API. Although there are some specific roles in GitHub, such as \textsf{maintainers} (promoted organization members who have a subset of privileges available to organization owners) and \textsf{collaborators} (outside contributors who have Read, Write, or Admin permissions), they are not always available via the API. To conduct a consistent analysis of all the targeted projects, we distinguish all the discussion participants based on GitHub terminology, as follows.

\begin{itemize}
    \item \textit{Member}: A participant belonging to an organization. After retrieving the list of organization members with the GitHub REST API v3, we investigate whether the participant is a member or not. If a project is owned by a personal developer, we consider the personal developer to be a member.
    \item \textit{Contributor}: A participant that is listed in the repository contributors of a project. We investigate whether the participant is not a member, but belongs to the repository contributors, which can also be obtained with the GitHub REST API v3.
    \item \textit{User}: Any participant who does not fit the above types.
\end{itemize}

\subsection{Survey (RQ2.2)}

To answer \textbf{RQ2.2}, we carry out a survey to get developer feedback on GitHub Discussions.
Following recent work~\citep{Robillard2020}, we structured our analysis of the survey answers based on the thematic analysis framework proposed by \cite{braun2006using}.
We process the results in two ways. For the quantitative results, we show the distribution statistics for each answer.
However, for the qualitative responses, we first identify whether or not the response is positive or negative against GitHub Discussions. 
Then in the results, we present representative quotes to support the themes that arise from the thematic analysis.
We ask developers the following three questions:
\begin{enumerate}
    \item \textit{What is your motivation to use GitHub Discussions for your project?}
    \item \textit{Do you find GitHub Discussions useful or redundant?}
    \item \textit{How does GitHub Discussions differ from the existing communication channels such as GitHub Issues, Pull Requests, or Stack Overflow?}
\end{enumerate}

As a follow-up to our early adopters survey, we sent a similar survey to the rest of GitHub users. At this point the GitHub Discussion function was open to all hosted projects.
We ask developers the following four questions:
\begin{enumerate}
    \item \textit{What do you think of GitHub Discussions?}
    \item \textit{Do you currently use an existing communication channel similar to GitHub Discussions in your software projects?}
    \item \textit{Do you have plans to use GitHub Discussions? Please, kindly add some reasons to support your decision}
    \item \textit{In your opinion, how does GitHub Discussions differ from submitting a GitHub Issue or Pull Request? Also, how does it compare to using external forums such as Stack Overflow?}
\end{enumerate}
Table \ref{tab:demographics} presents an overview of the demographics of our survey participants.
The main survey was distributed to all 98 projects we identified as having GitHub Discussions activated.
On August 11 and 12, 2020, we posted our survey directly to the discussion channel in each project, yielding 19 responses.
Similar to the first survey, the follow-up survey was distributed to 2,000 randomly selected npm developers that hosted their projects on GitHub. 
By May 2021, the survey had yielded 52 responses.
More than half of respondents are core contributors, such as maintainers or lead developers.
Programming experience varies from less than 5 years to 30 and above.

\begin{table}[t]
\centering
\caption{Demographics of survey participants}
\label{tab:demographics}
\begin{tabular}{llcc}
\textbf & & first survey (n=19) & follow-up survey (n=52)\\
\toprule
\textbf{role} & core contributor & 10 & -\\ 
& regular contributor & 8 &-\\ 
& newcomer & 1 &-\\ 
\midrule
\textbf{gender} & male & 16 & 48\\ 
& prefer not to  & 3 & 4\\ 
&say/non-binary\\
&female &-&-\\
\midrule
\textbf{programming} & less than 5 & 2 & -\\ 
\textbf{years} & 5-9 & 6 & 10\\ 
& 10-19 & 6 & 14\\ 
& 20-29 & 3 & 13\\ 
& 30 and above & 2 &8\\ 
\bottomrule
\end{tabular}
\end{table}

\subsection{Sentiment Analysis (RQ3.2)}

To answer \textbf{RQ3.2}, we analyse and compare the sentiment polarity of posts in GitHub Discussions and in Stack Overflow. Our focus on sentiment is motivated by repeated mentions to Stack Overflow being `anti-discussion' on Hacker News\footnote{\url{https://news.ycombinator.com/item?id=22388639}} and related work which has identified unfriendly comments on Stack Overflow~\citep{Cleary2013}. Being explicitly `pro-discussion', we speculate that GitHub Discussions could be a forum with comparatively positive sentiment.
To this end, we use the collected 32,714 Discussion posts.
In parallel, we collect questions and answers from Stack Overflow threads related to the targeted projects. We manually searched Stack Overflow tags for repository and owner names, resulting in a set of 79 tags (82\% of project coverage) as some projects did not have specific tags on Stack Overflow. The list of tags is included in our replication package in Zenodo. 

Using the Stack Exchange Data Explorer,\footnote{\url{https://data.stackexchange.com/stackoverflow/query/new}} we collected 99,918 posts (questions with the tags and their answers) created between January and July, 2020. We preprocessed all the posts to remove HTML tags and code snippets using the Beautiful Soup library.\footnote{\url{https://pypi.org/project/beautifulsoup4/}}
To extract the sentiment conveyed by the posts in the GitHub Discussions and in Stack Overflow questions and answers, we use two sentiment analysis tools specifically tuned for the software engineering domain, namely SentiStrength-SE and Senti4SD, 
as they have been demonstrated to be robust enough in either within- or cross-platform settings~\citep{10.1145/3196398.3196403}.

\textbf{Senti4SD}~\citep{CalefatoEMSE18} is trained by leveraging supervised machine learning on a manally annotated gold standard of $\sim$4K questions, answers, and comments from Stack Overflow. It exploits a suite of features including bag of words, sentiment lexicons, and semantic features based on word embedding. 
For this analysis, we used the Python version of Senti4SD~\citep{EmtkSemotion2019}. Since our goal is to label data from Stack Overflow and GitHub, we use Senti4SD in two different conditions: as off-the-shelf tool, leveraging the classification model originally trained on Stack Overflow, and by retraining it on the GitHub gold standard dataset provided by Novielli et al.~\citep{10.1145/3196398.3196403}, in line with their recommendations to retrain supervised tools with data from the target platform.
Senti4SD outputs a polarity label in $\{positive,negative,neutral\}$, consistently with the gold labels in the training sets used for training. 

\textbf{SentiStrength-SE}~\citep{Islam17} implements a set of heuristics leveraging sentiment lexicons, that is, large collections of words annotated with their prior polarity, i.e., the positive or negative orientation of the word. The overall sentiment of a text is computed based on the prior polarity of the words composing it, under the assumption that words with negative prior polarity convey negative sentiment, and vice versa. SentiStrength-SE is built upon the API of the general-purpose tool SentiStrength~\citep{Thelwall2010}. It leverages a manually adjusted version of the SentiStrength lexicon and implements \textit{ad hoc} heuristics to correct the misclassifications due to domain-specific semantics of terms in the lexicon. 
SentiStrength-SE assigns an integer value between $1$ and $5$ for the positivity of a text, $p$ and, analogously, between $-1$ and $-5$ for the negativity $n$.
We map these scores 
to polarity values in $\{positive, negative, neutral, mixed\}$. Specifically, a text is considered positive when $p + n > 0$, negative when $p + n < 0$, and neutral if $p = 1$ and $n  = -1$. Texts with a score of $p = -n$ are considered mixed.

\begin{table}[htb]
\caption{Majority voting criterion for polarity label assignment using the polarity labels from SentiStrength-SE and the two versions of Senti4SD.}
\label{tab:sentimentExamples}
\resizebox{\textwidth}{!}{%
\begin{tabular}{l|ccc|cc|c}
\hline
 & \multicolumn{3}{c|}{\textbf{SentiStrength-SE}} & \multicolumn{2}{c|}{\textbf{Senti4SD}} &  \\ \cline{2-6}
\multirow{-2}{*}{\textbf{Input text}} & \textbf{\begin{tabular}[c]{@{}c@{}}Positive \\ score\end{tabular}} & \textbf{\begin{tabular}[c]{@{}c@{}}Negative \\ Score\end{tabular}} & \textbf{\begin{tabular}[c]{@{}c@{}}Polarity \\ label\end{tabular}} & \textbf{\begin{tabular}[c]{@{}c@{}}Trained \\ on SO\end{tabular}} & \textbf{\begin{tabular}[c]{@{}c@{}}Trained \\ on GH\end{tabular}} & \multirow{-2}{*}{\textbf{Final label}} \\ \hline
\textit{``Which version of AdonisJS?''} & 1 & -1 & neutral & neutral & neutral & neutral \\
\rowcolor[HTML]{EFEFEF} 
\textit{``Closing discussion is low quality''} & 1 & -1 & neutral & negative & negative & negative \\
\textit{``Glad, you found it :)''} & 2 & -1 & positive & positive & positive & positive \\
\rowcolor[HTML]{EFEFEF} 
\textit{\begin{tabular}[c]{@{}l@{}}``You can also force the \textless{}CODE\textgreater in your configuration \textless{}CODE\textgreater{}. \\ Thanks! You shouldn't pass an anonymous class. \\ Yes, this is a very bad example! Only for demo, no one use it :D'' \end{tabular}} & 2 & -3 & mixed & negative & negative & mixed \\ \hline
\end{tabular}%
}
\end{table}

We assign the final polarity label for each text in our datasets based on majority agreement between the output of SentiStrength-SE and the two classification models of Senti4SD. Specifically, positive, neutral, and negative labels are assigned based on majority voting, that is when at least two tools agree on a given polarity class (see the first three examples in Table~\ref{tab:sentimentExamples}). Mixed cases are those classified as such by Sentistrength-SE, which issues to separate predictions for positive and negative polarity (see last example in the table).

\section{Results}
\label{sec:findings}

In the following, we answer our research questions with the data collected according to the research methodology described above.
Each sub-question is presented in a dedicated section, ending with a summary of the corresponding results.

\subsection{Discussion Categories (RQ1.1)}
\label{sec:discussionCategories}

\begin{table}[t]
\centering
\caption{Frequency of discussion categories in our samples}
\label{tab:category}
\begin{tabular}{lr@{}rr@{}r}
\toprule
\textbf{category} & \multicolumn{2}{c}{\textbf{from issues}} & \multicolumn{2}{c}{\textbf{not from issues}} \\
\midrule
Errors & 102 & (37\%) & 77 & (21\%) \\
Discrepancy & 25 & (9\%) & 134 & (37\%) \\
Review & 40 & (14\%) & 80 & (22\%) \\
Conceptual & 45 & (16\%) & 25 & (7\%) \\
Usage & 27 & (10\%) & 12 & (3\%) \\
Learning & 12 & (4\%) & 11 & (3\%) \\
Versions & 11 & (4\%) & 4 & (1\%) \\
Plans & 8 & (3\%) & 5 & (1\%) \\
Announcement & 2 & (1\%) & 0 & (0\%) \\
Information & 1 & (0\%) & 2 & (1\%) \\
Recruitment & 0 & (0\%) & 2 & (1\%) \\
Other & 5 & (2\%) & 5 & (1\%) \\
404 & 0 & (0\%) & 4 & (1\%) \\
\midrule
\textbf{sum} & \textbf{278} & \textbf{(100\%)} & \textbf{361} & \textbf{(100\%)} \\ \bottomrule
\end{tabular}
\end{table}

To answer our first research question (What are GitHub Discussions about?), we manually annotated a representative sample of discussion threads.

Table~\ref{tab:category} shows the result of our annotation. A prevalence of Errors (37\%) is observed in the discussions originating from previous Issues (`from issues' in the table), which denotes the goal of asking for support in fixing errors or understanding exceptions. As these discussions had been converted from Issues, it is revealed that developers prefer to manage errors on GitHub Discussions until the underlying bugs have been identified.
Errors are reported only in 21\% of posts originally started as discussion (`not from issues' in the table). Conceptual and Usage questions are also more frequent in the `from issues' group.
Conversely, we observe a prevalence of Discrepancy (37\%) in the `not from issues' sample, compared to 9\% observed for the `from issues' group.
This might be an indication of the more open nature of the Discussions, with respect to Issues or other Q\&A platforms, such as Stack Overflow. In fact, Discrepancy refers to the kind of questions that are not well-suited for Issues due to their high-level of abstraction and due to the fact that the authors of such posts have no clue how to solve the problems they describe.
Indeed, this is the kind of questions that would not be appreciated on Stack Overflow, whose guidelines explicitly recommend information seekers to provide evidence of previous attempts to solve the problem.\footnote{`Writing the perfect question' by Jon Skeet at \url{https://codeblog.jonskeet.uk/2010/08/29/writing-the-perfect-question/} linked in the official Stack Overflow guidelines at \url{https://stackoverflow.com/help/how-to-ask}} For example, Conceptual is considered to be a category related to such evidence, and is less common in `not from issue' (7\%) than in `from issue' (16\%).
Conversely, Review is more frequent in `not from issue' (22\%) than in previous Issues (14\%).


We compare the distribution of categories for GitHub Discussions with the distribution reported by~\cite{Beyer2020} for Stack Overflow. In their study, they performed automated categorization of Stack Overflow questions with the above-mentioned taxonomy and obtained the following distribution of question categories:
Errors (19\%), Discrepancy (26\%), Review (7\%), Conceptual (22\%), API Usage (36\%), Learning (2\%), and API change (2\%).
Compared to such distribution, there are less Conceptual and Usage, and more Discrepancy and Review discussions in GitHub Discussions `not from issues'. We conjecture that this is because there is no strict requirement for providing evidence (leading to more Discrepancy) and because of the increased connectivity to source code (leading to more Review) for GitHub Discussions.
Although the percentages are small, there are some discussion categories that do not appear in Stack Overflow, such as Plans, Announcement, Information, and Recruitment in GitHub Discussions, which indicate that GitHub Discussions is more open channels for communication and a community knowledge base.

\begin{tcolorbox}
\textbf{Summary}: Errors, Discrepancy, and Review are prevalent discussion categories in GitHub Discussions, whereas various other categories, such as Announcement, Information, and Recruitment also exist.
\end{tcolorbox}

\subsection{Discussion Participants (RQ1.2)}

\begin{table}[t]
\centering
\caption{Discussion patterns by questioners}
\label{tab:response1}
\begin{tabular}{lr@{}rr@{}rr@{}r}
\toprule
 & \multicolumn{2}{c}{\textbf{user}} & \multicolumn{2}{c}{\textbf{contributor}} & \multicolumn{2}{c}{\textbf{member}} \\
\midrule
has selected answer & 2,550 & (39\%) & 183 & (44\%) & 33 & (23\%) \\
no selected answer & 2,464 & (39\%) & 165 & (40\%) & 79 & (55\%) \\
no response & 1,542 & (24\%) & 68 & (16\%) & 32 & (22\%) \\
\midrule
\textbf{sum} & \textbf{6,556} & \textbf{(100\%)} & \textbf{416} & \textbf{(100\%)} & \textbf{144} & \textbf{(100\%)} \\
\bottomrule
\end{tabular}
\end{table}

To investigate our second research question (Who contributes to GitHub Discussions?), we distinguish between the roles of Member, Contributor, and User.

Table~\ref{tab:response1} shows the discussion patterns divided by questioners who started discussions. From all 7,116 discussions, the majority have been started by Users, followed by Contributors and Members. For discussions started by all three types, more than three quarters have responses. Similar to Stack Overflow, GitHub Discussions allows users to select answers in discussions. However, as seen in Section~\ref{sec:discussionCategories}, not all discussion categories require to select single answers, which could be seen in the large percentages of discussions without selected answers (39-55\%).

\begin{table}[t]
\centering
\caption{Respondents of selected answers by questioners}
\label{tab:response2}
\begin{tabular}{lr@{}rr@{}rr@{}r}
\toprule
 & \multicolumn{2}{c}{\textbf{user}} & \multicolumn{2}{c}{\textbf{contributor}} & \multicolumn{2}{c}{\textbf{member}} \\
\midrule
answered by member & 1,162 & (46\%) & 106 & (58\%) & 19 & (58\%) \\
answered by contributor & 578 & (23\%) & 72 & (39\%) & 9 & (27\%) \\
answered by user & 810 & (32\%) & 5 & (1\%) & 5 & (15\%) \\
\midrule
\textbf{sum} & \textbf{2,550} & \textbf{(100\%)} & \textbf{183} & \textbf{(100\%)} & \textbf{33} & \textbf{(100\%)} \\
\bottomrule
\end{tabular}
\end{table}

\begin{table}[t]
\centering
\caption{Respondents for discussions without answers by questioners}
\label{tab:response3}
\begin{tabular}{lr@{}rr@{}rr@{}r}
\toprule
 & \multicolumn{2}{c}{\textbf{user}} & \multicolumn{2}{c}{\textbf{contributor}} & \multicolumn{2}{c}{\textbf{member}} \\
\midrule
member responded & 1,029 & (42\%) & 82 & (50\%) & 69 & (87\%) \\
contributor responded & 554 & (22\%) & 72 & (44\%) & 4 & (5\%) \\
only user responded & 881 & (36\%) & 11 & (7\%) & 6 & (8\%) \\
\midrule
\textbf{sum} & \textbf{2,464} & \textbf{(100\%)} & \textbf{165} & \textbf{(100\%)} & \textbf{79} & \textbf{(100\%)} \\
\bottomrule
\end{tabular}
\end{table}

For discussions that have selected answers, Table~\ref{tab:response2} presents the distributions of answerers by questioners. For all questioners, answers by Members are most frequently selected.
Table~\ref{tab:response3} shows the participation patterns in discussions by questioners, for discussions without selected answers. We divide discussions into three types, that is, discussions in which at least one Member responded, discussions for which no Member was involved but Contributors responded, or discussions in which only Users responded. Similar to Table~\ref{tab:response2}, Member-responded discussions are most frequent.
These results indicate that Members actively participated in GitHub Discussions.

In addition, we investigated the response time of each Member, Contributor, and User. The response time was measured from the time a question was posted until a different person than the one who initially asked the question commented on it. The median response time was 4.4 hours for Members, 4.1 hours for Contributors, and 8.5 hours for Users. When the Mann-Whitney U test was used to compare the differences in response times, there was no statistically significant difference between Members and Contributors, but there were statistically significant differences between Members and Users, and between Contributors and Users. We can see that comments from Members and Contributors were obtained earlier than comments from other Users.

\begin{tcolorbox}
\textbf{Summary}: Most of the discussions are initiated by Users, and Members were actively responding to them on GitHub Discussions. Also, Members and Contributors responded earlier than other Users.
\end{tcolorbox}

\subsection{Project Characteristics (RQ1.3)}
\label{ssec:1.3}

\begin{table}[t]
\centering
\caption{Domains of Target Projects}
\label{tab:domain}
\begin{tabular}{lrr}
\toprule
\textbf{domain} & \textbf{\# projects} & \textbf{\%} \\
\midrule
Web libraries and frameworks & 37 & 38\% \\
Software tools & 32 & 33\% \\
Application software & 13 & 13\% \\
Non-web libraries and frameworks & 8 & 8\% \\
System software & 8 & 8\% \\
\midrule
\textbf{sum} & \textbf{98} & \textbf{100\%} \\
\bottomrule
\end{tabular}
\end{table}

To answer our next research question (What is the relationship between project characteristics and GitHub Discussion adoption?), we employed a qualitative analysis to reveal software domains of the targeted projects and analyzed the relationship of core developer involvement and discussion activities.

\begin{figure*}
  \centering
  \includegraphics[width=\linewidth]{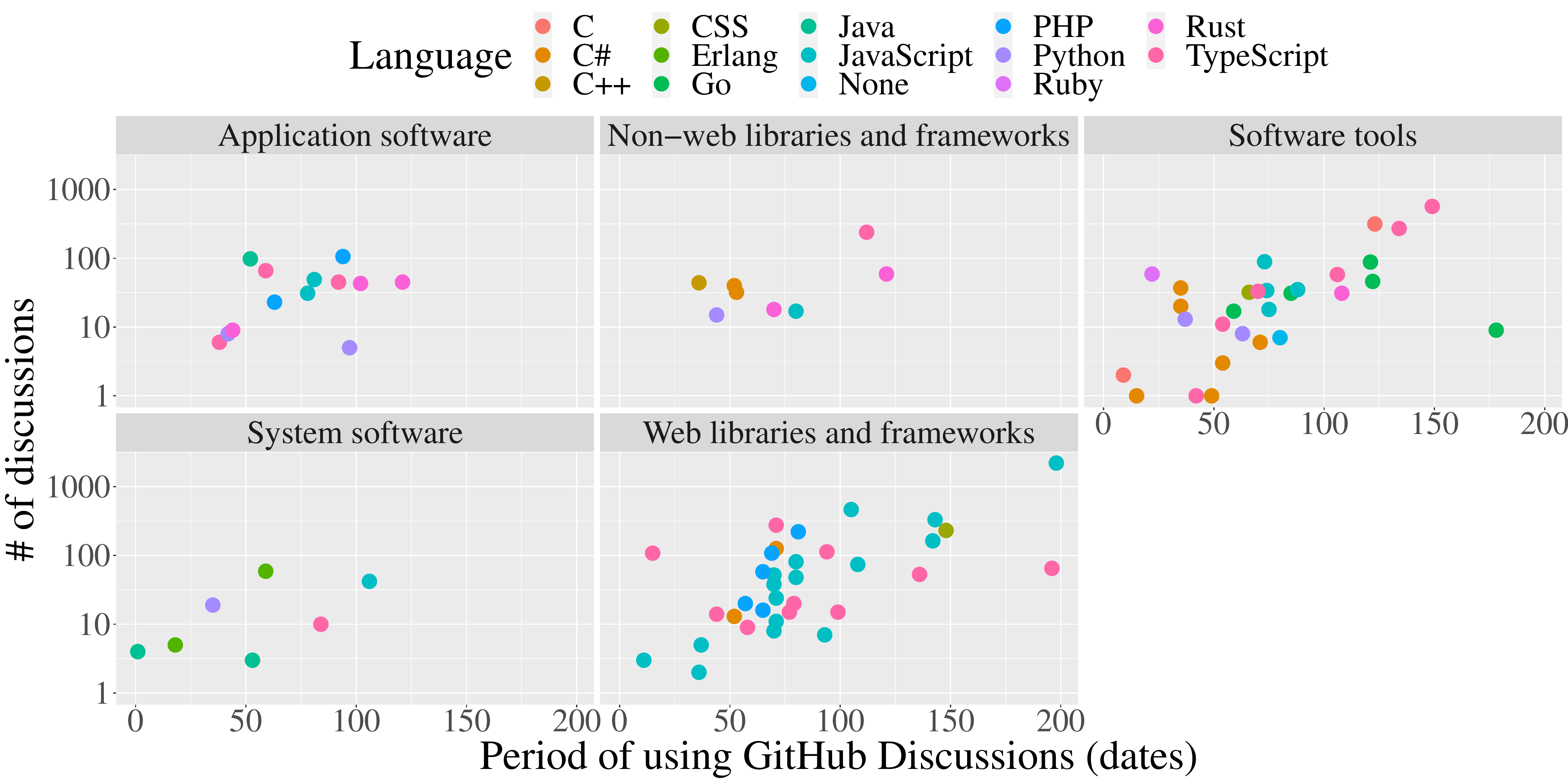}
  \caption{The number of discussions (log scale) by the period of using GitHub Discussions.}
  \label{fig:distribution}
\end{figure*}

The result of the annotation of the software domain is reported in Table~\ref{tab:domain} and indicates a prevalence of \textit{Web libraries and framework}, immediately followed by \textit{Software tools} and \textit{Application software}.
A previous study that investigated the top 2,500 repositories by number of stargazers on GitHub reported that there were many \textit{Non-web libraries and frameworks}, followed by \textit{Web libraries and frameworks}. In comparison, there were not many Non-web libraries and frameworks in early adopters of GitHub Discussions.
%
%
Figure~\ref{fig:distribution} presents the distributions of the targeted projects based on the periods using GitHub Discussions and the total number of discussions. The periods were derived by identifying the initial discussions that were not converted from existing Issues. Seven projects are not shown since they had no discussions or had only discussions converted from issues.
Programming languages of the projects retrieved using the GitHub REST API are shown with different colors, revealing that the studied projects use various different programming languages.
As expected, there is a correlation between the number of discussions and the period of using GitHub Discussions. However, some projects have relatively few discussions considering their discussion periods.

\begin{figure}[t]
  \centering
  \includegraphics[width=\linewidth]{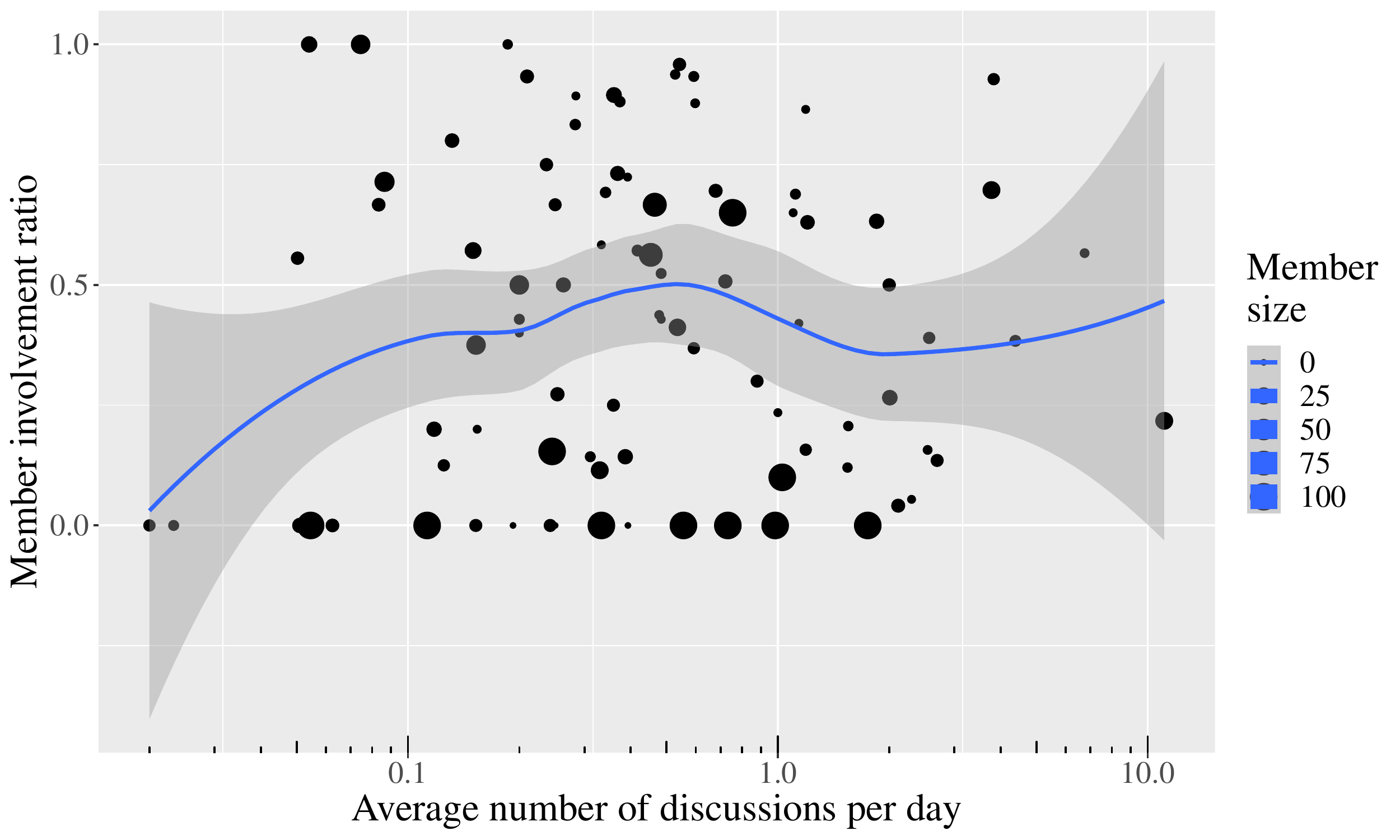}
  \caption{Membership involvement ratio by average number of discussions per day (log scale). Point size represents the number of Members in a project.}
  \label{fig:member-ratio}
\end{figure}

To understand the characteristic of discussion participation and discussion activities, Figure~\ref{fig:member-ratio} illustrates the relations of the average number of discussions per day and the ratio of Member involvement, which is the ratio of
discussions that contain at least one post added by a Member to the total number of discussions. This is measured based on the posts that were included in the discussion at the time of collection.
A regression line is plotted using LOESS smoothing, which fits multiple regressions in a local neighborhood~\citep{10.1007/978-3-642-48425-4_2}. The shaded area indicates the confidence intervals. We observe that the Member involvement ratio is increasing until 0.5 in the average number of discussions per day, which could be considered as less active, and decreases and disperses for more than 0.5 average discussions per day.
This suggests that high Member involvement encouraged active use of GitHub Discussions, and, for projects with many discussions per day, Contributors and Users also contributed to answering questions.

\begin{tcolorbox}
\textbf{Summary}: The early adopters of GitHub Discussions include projects developed in a variety of programming languages, especially in the domains of ``Web libraries and frameworks'' and ``Software tools''.
Among the projects using GitHub Discussions, there is a positive relationship between Member involvement and discussion frequency for projects with less active discussions.
\end{tcolorbox}


\subsection{Initial Discussions (RQ2.1)}
\label{ssec:2.1}

\begin{table}[t]
\centering
\caption{Reasons for using GitHub Discussions mentioned in initial discussions. Multiple reasons per project possible.}
\label{tab:initial}
\begin{tabular}{lr}
\toprule
\textbf{reason} & \textbf{\# projects} \\
\midrule
question-answering & 16 \\
idea sharing & 12 \\
community engagement & 12 \\
decluttering issue tracker & 3 \\
information resource building & 2 \\
feature requests & 2 \\
thank-you & 1 \\
testing the feature & 16 \\
not meta & 51 \\
\bottomrule
\end{tabular}
\end{table}

To investigate why projects adopt GitHub Discussions, we qualitatively analyzed the first discussion thread of each project in our data. This methodology was motivated by our anecdotal observation that the first discussion thread was often used to discuss the Discussions feature and how to best use it, rather than project-specific issues.

Table~\ref{tab:initial} shows the distribution of the codes across the first discussion threads we annotated. Note that one thread might mention multiple reasons for using GitHub Discussions. About half of the first discussion threads were not used for meta discussions (`not meta' in Table~\ref{tab:initial}), but were instead used to discuss project-specific issues. For example, the first discussion thread of the PrefectHQ/prefect project\footnote{\url{https://github.com/PrefectHQ/prefect/discussions/2860}} asks for help with a particular error message. The remaining threads mentioned various reasons for using the GitHub Discussions feature, in particular question-answering, idea sharing, and community engagement. To further understand developers' rationale, we conducted a survey, see next section.

\begin{tcolorbox}
\textbf{Summary}: For projects which used their first discussion thread to discuss how to use the new feature, question-answering, idea sharing, and community engagement were among the top reasons mentioned.
\end{tcolorbox}

\subsection{Developer Survey (RQ2.2)}
\label{ssec:survey}

To understand how software developers perceive the GitHub Discussions feature, we conducted a survey that 1) targeted the early adopters of GitHub discussions and 2) targeted the rest of GitHub projects.

\textbf{What is your motivation to use the discussion channel for your project? (19 responses)}:
Participants provided a range of answers that revolved around being a community medium to act as a general forum, and not necessarily about Issues. One participant responded:
\begin{quote}
    \textit{It's the perfect place to discuss feature request and other communication forms (like questions) that don't really fit in Issues.}
\end{quote}
Four participants mentioned that they liked that discussions were closer to the code (within the GitHub Platform), as generic questions:
\begin{quote}
    \textit{We wanted to offload the constant stream of questions about the project to someplace other than the issue tracker, but still linked to GitHub.}
\end{quote}

\textbf{Do you find GitHub Discussions useful or redundant? (19 responses)}:
15 out of the 19 participants were very positive about the usefulness of GitHub Discussions. One participant responded that it was useful to store drawn-out discussions for the community:
\begin{quote}
    \textit{Useful: allows us to have long, drawn-out discussions that don't necessarily have to have a conclusion. 
    }
\end{quote}
On the other hand, there were three participants that were not as positive towards GitHub Discussions.
One participant 
%
%
found the discussions difficult to differentiate from Issues and Pull Requests.
\begin{quote}
    \textit{Redundant: people often don't know whether to use a discussion or an issue or a PR. Many discussions end up being opened that are duplicates of issues (and vice versa). In addition, speaking entirely from impressions, it seems that discussion posters do not seem to hold themselves to the same standards as issue posters. 
    }
\end{quote}
\begin{table}[t]
\centering
\caption{Response if respondents use a channel similar to GitHub Discussions}
\label{tab:exist}
\begin{tabular}{lr@{}rl}
\toprule
Using Similar Channel & \multicolumn{2}{c}{\# Responses} & \\
\midrule
Yes                   & 19                 & & \mybar{0.365}                      \\

No                   & 26                    &   & \mybar{0.5}                       \\

Other              & 7                  &      &  \mybar{0.07}                  \\
\midrule
Total                      & 52                 &               \\   
\bottomrule
\end{tabular}
\end{table}

\begin{table}[]
\centering
\caption{Response if respondents consider to use GitHub Discussions}
\label{tab:plans}
\begin{tabular}{lr@{}rl}
\toprule
Plans to use GitHub Disussions & \multicolumn{2}{c}{\# Responses} & \\
\midrule
Yes                   & 23                &  & \mybar{0.23}                      \\

No                   & 21                    &   & \mybar{0.21}                       \\

Other              & 8                    &     &  \mybar{0.0033}                  \\
\midrule
Total                      & 52                &                \\   
\bottomrule
\end{tabular}
\end{table}

\textbf{How does GitHub Discussions differ from the existing communication channels such as submitting a GitHub Issue or Pull Request? Also, how does it compare to other external forums such as Stack Overflow? (16 responses}):
One participant 
noted that:
\begin{quote}\textit{Way more appropriate for questions and other topics that are not actionable for the engineering team in any way. GitHub issues and PRs are used by our engineering team as the foundation for their process. 
}    
\end{quote}
Another participant pointed out the accessibility compared to Stack Overflow:
\begin{quote}
\textit{discussions offer a place where we can discuss ideas that aren't necessarily actionable points. Discussions is very similar to StackOverflow in it's approach, but it's way more accessible as it's scoped to the project in question and doesn't have the barrier of entry (due to the somewhat hostile community) that StackOverflow does.}
\end{quote}
Overall, when comparing to Stack Overflow, there was no uniform opinion as we received both positive and negative responses.
One participant noted that the usefulness of a discussion medium also depends on the nature of the project:

\begin{quote}
    \textit{For our project, the discussion is a much better forum than StackOverflow (SO) because we are not a very well-known project and have some strange quirks in our configuration rules. For more widespread projects, I would likely prefer SO simply because it has more reach.}
\end{quote}

For the rest of GitHub projects that have not adopted GitHub Discussions, they provided the following responses to our survey:

\textbf{What do you think of GitHub Discussions?  (52 responses}):
All responses were positive, stating that GitHub Discussions was different from a pull request or issue. For instance:
\begin{quote}
    \textit{I like the nested threads. Regular github issues don't have nested threads so the conversations get a bit shuffled.}
\end{quote}

\textbf{Do you currently use an existing communication channel similar to GitHub Discussions in your software projects? (52 responses)}:
Table \ref{tab:exist} show the results of the follow-up survey sent to the rest of GitHub projects. 
The results were mixed, with 26 out of the 52 respondents claiming that they did not use such a similar channel for their communication. On the other hand, 19 respondents claimed they had an existing communication channel like GitHub Discussions. 

\textbf{Do you have plans to use GitHub Discussions? (52 responses)}:
Similar the previous question, Table \ref{tab:plans} shows a mixed response to whether or not the participants had plans to adopt GitHub Discussions into their projects. Still a bit over 50\% (23 participants) were willing to use GitHub Discussions.

\textbf{In your opinion, how does GitHub Discussions differ from submitting a GitHub Issue or Pull Request? Also, how does it compare to using external forums such as Stack Overflow? (46 responses)}:
Different to the early adopters responses, we find that there were more respondents that were positive in regards to the difference between GitHub Discussions and a GitHub Issue or a Pull Request. 
Applying a thematic analysis approach, we were able to extract the themes of how the respondents expressed themselves.
Overall, we found 31 responses to be more positive when explaining the differences of GitHub Discussions. Most discussed the structure and how there was a need for a discussions thread that fell outside of the development pipelines. For instance, 
\begin{quote}
    \textit{Most questions and discussion should not be posted as Issues. However, an open source project can generate a lot of valid discussion. This is where GitHub Discussions will shine and for the reasons mentioned above: open-source communities are oftentimes overwhelmed by requests that are not issues.}
\end{quote}
On the other hand, there were also responses that were not as positive. We found six respondents who raised concerns. For instance, 
\begin{quote}
    \textit{GH Issues (and pulls) are still the preferred way to tell the owner/maintainer that there is a bug/feature request/question. They are enabled on all repos by default whereas Discussions need to be enabled. StackOverflow has a wider community would would better be able to answer any question you throw at them, whereas GH Discussions are better suited to questions which are very specific to the project you are asking them in.}
\end{quote}
Finally, we found 9 responses that were either unable to tell the difference, or they were neutral with their stance for GitHub Discussions. For example, 
\begin{quote}
    \textit{Compared to Stack Overflow: sounds very similar - we will see which system will win :)}
\end{quote}

\begin{tcolorbox}
\textbf{Summary}: 
Developers are able to distinguish between question-answering and engineering tasks and react positively to the accessibility and integration of  GitHub Discussion with other GitHub features, but they also point to the problem of topic duplication between Discussions and Issues.
For developers that had not adopted GitHub Discussions, although there were still mixed reactions to its adoption, overall the feedback was positive, often with a wait-and-see approach.
\end{tcolorbox}


\subsection{Impact on Other Channels (RQ3.1)}
\label{ssec:3.1}

\begin{table}[t]
\centering
\caption{Why do links to Issues or Pull Requests exist? Result on a sample of 231 links.}
\label{tab:links}
\begin{tabular}{llr}
\toprule
\textbf{link} & \multicolumn{2}{c}{\textbf{reason of link appearance}} \\
\midrule
& newly open & 42 \\
issue & deepen discussion & 11 \\
& reference & 90 \\
\midrule
& newly open & 37 \\
pull requests & deepen discussion & 5 \\
& reference & 46 \\
\midrule
\textbf{sum} & & \textbf{231}\\
\bottomrule
\end{tabular}
\end{table}

To investigate the impact of GitHub Discussions on other channels, we conducted a qualitative analysis of a representative sample of links to Issues and Pull Requests from Discussions. We only considered links to artefacts that had been opened or closed after the corresponding Discussion started.

Table~\ref{tab:links} shows the distribution of reasons for links to Issues and Pull Requests from GitHub Discussions. Out of 231 linked Issues and Pull Requests in our sample, 79 (34\%) were created as a result of a GitHub Discussion. In many other cases, links were provided for reference.

\begin{tcolorbox}
\textbf{Summary}: GitHub Discussions does not exist in isolation, but can play a crucial role in moving development forward by triggering new Pull Requests or Issues.
\end{tcolorbox}

\subsection{Sentiment Characteristics (RQ3.2)}
\label{ssec:3.2}


\begin{table}[t]
\centering
\caption{Sentiment analysis of all posts and accepted answers in the GitHub Discussions and Stack Overflow Q\&A threads.}
\label{tab:sentiment}
\begin{tabular}{lrrrr}
\toprule
 & \multicolumn{2}{l}{\textbf{GH Discussions posts}} & \multicolumn{2}{l}{\textbf{Stack Overflow posts}} \\
\textbf{polarity} & \textit{all} & \textit{accepted} & \textit{all} & \textit{accepted} \\
\midrule
positive & 12\% & 13\% & 3\% & 4\% \\
negative & 6\% & 4\% & 6\% & 3\% \\
neutral & 73\% & 78\% & 82\% & 91\% \\
mixed & 9\% & 6\% & 9\% & 3\% \\
\midrule
\textbf{Overall items} & 32,714 & 4,317 & 99,918 & 19,227 \\
\bottomrule
\end{tabular}%
\end{table}

To answer our final research question (What are sentiment differences compared to Stack Overflow posts?), we analysed and compared the sentiment polarity of posts in GitHub Discussions and Stack Overflow.
The voting label was assigned either with full agreement (the three tools agree) in 64\% (GitHub) and 61\% (Stack Overflow) of cases or by majority voting in 27\% and 30\% of cases (2 over 3 agree),  indicating a substantial observed agreement between the tools. Similarly to Stack Overflow, in GitHub Discussions, an answer can also be flagged as accepted. As such, we complement our analysis by also investigating whether there is any difference in the accepted answers between GitHub Discussions and Stack Overflow.  The result of the sentiment analysis is reported in Table~\ref{tab:sentiment}.

In line with previous work~\citep{CalefatoIST18, Murgia2014} we observe that the majority of posts is neutral, i.e. they do not convey any emotions or opinions, which is reasonable, given the technical nature of Discussions. The amount of both mixed and negative posts in the two samples is comparable. As for positive sentiment, we observe that it is present in 12\% of GitHub Discussions while it appears in 3\% only of Stack Overflow posts, for which we observe also a higher percentage of neutral cases. This might be due to the Stack Overflow community adhering to a more neutral communication style, in line with official recommendations. A recent empirical study demonstrated how expressing either positive or negative emotions is associated with a lower probability of obtaining a useful answer~\citep{CalefatoIST18}, which further supports our conjecture.
In both datasets, accepted answers appear more neutral, as a consequence of the less negative and mixed sentiment. We also observe a small increment in positive sentiment. It is the case, for example, of sentences expressing either wish (e.g., `See the code below! Happy 2020!'), politeness expressions (e.g., `I hope this solves your issue') or positive sentiment in formulating requests (e.g., `It would be nice if we can convert that issue into a discussion \smiley{}').
To conclude, the differences in the sentiment observed for the two samples are mainly due to GitHub Discussions being more positive than SO posts. The results of chi-squared test applied to the contingency table show that the differences are statistically significant ($p < .05$) but the effect size is negligible (Cramer's V = $.17$)~\citep{cohen1988spa}.

Previous results of qualitative analysis of emotions in Stack Overflow suggest that a polarity-based operationalization of affect might not be adequate to capture the nuances in the attitude conveyed by negative sentiment~\citep{CalefatoIST18}.  
In fact, the presence of negative sentiment does not necessarily indicate the use of abusive language or toxic interactions~\citep{Gachechiladze17}. Being able to identify rude, aggressive comments among negative ones is crucial, as toxicity might impair effective collaboration and prevent people from contributing to software projects~\citep{Steinmacher:2014}. 

\begin{figure}[t]
  \centering
  \includegraphics[width=.8\linewidth]{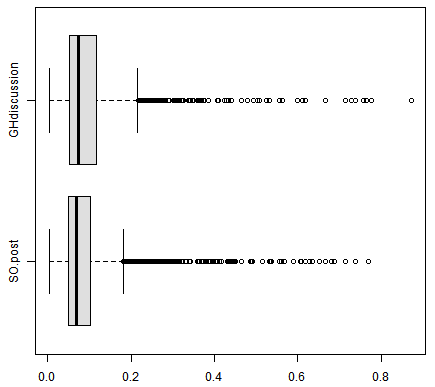}
  \caption{Toxicity in negative posts on GitHub Discussions and Stack Overflow.}
  \label{fig:toxicity}
\end{figure}

In line with this view, we performed a follow-up analysis to investigate the toxicity of 1,962 posts from GitHub Discussions and 5,635 questions and answers from Stack Overflow that were classified as negative. Consistently with previous research~\citep{Raman2020StressAB}, we computed the toxicity scores for both samples of negative posts, using the Google Perspective API,\footnote{\url{www.perspectiveapi.com}} specifically designed to identify abusive language and harassment in online conversations. The Perspective API provides, for a given text in input, a toxicity score in $[0,1]$ representing the probability that  a reader would perceive the text as rude or disrespectful, leading him/her to abandon the conversation. Results are reported in Figure~\ref{fig:toxicity} and show that toxic posts are rare in line with previous evidence provided by \cite{Raman2020StressAB}, who analyzed the toxicity in open source discussions.

\begin{tcolorbox}
\textbf{Summary}: The majority of Discussion posts is neutral. Toxicity and abusive language are rarely detected. The sentiment is comparable to the sentiment of Stack Overflow posts except for positive sentiment, which is more frequent in GitHub Discussions.  
\end{tcolorbox}

\section{Discussion}

Our findings can be summarized into the following recommendations for practitioners.
\begin{itemize}
    \item 
    \textit{Set guidelines for participating in discussions.}
    As found in the response from developers in Section~\ref{ssec:survey}, users sometimes find it difficult to choose an appropriate channel from Discussions, Issues, and Pull Requests. It should be beneficial for community members to set a policy of using these channels depending on the project-specific development processes.
    \item 
    \textit{Encourage core developers to participate in discussions.}
    It is evident that the involvement of core developers is crucial in the successful use of GitHub Discussions as seen in Section~\ref{ssec:1.3}.
    \item
    \textit{Prepare for newcomers.}
    GitHub Discussions can be a gateway for newcomers to participate by creating discussions specifically for them~\citep{8254320}. Not only does a project present tasks that are easy for newcomers to start, but it can also provide a place for mentoring and effective communication between mentors and newcomers, potentially solving the mentoring challenges pointed out in a recent study~\citep{10.1145/3412569.3412571}.
\end{itemize}

Besides, our analysis has shown that Stack Overflow discussions are more neutral compared to GitHub Discussions and also less positive (see Section~\ref{ssec:3.2}).
Software projects might use this as an opportunity to establish GitHub Discussions as a feedback channel that explicitly encourages users to utter their appreciation for a certain feature or a response they received in a discussion.
This would also help to distinguish GitHub Discussions from discussions on Stack Overflow, where, e.g., thank you messages are discouraged.\footnote{\url{https://codeblog.jonskeet.uk/2010/08/29/writing-the-perfect-question/}}

From our exploratory study of early adoption of GitHub Discussions, we see the following avenues for future research:
\begin{itemize}
    \item 
    \textit{Tool support for effective discussions.}
    As developers mentioned (Section~\ref{ssec:survey}), duplicate posts in Discussions and Issues could be troublesome.
    Automated support using bots like \citep{Abdellatif2020} is promising.
    \item 
    \textit{Further studies of error/issue handling.}
    As we saw in Section~\ref{sec:discussionCategories}, more than a third of the issues converted to discussions were ``errors''. This can be attributed to the fact that some projects prefer to have discussions on how to deal with errors, unexpected behavior, etc., and to determine if errors are bugs, in discussions rather than in issues.
    As seen in Section~\ref{ssec:3.1}, there are issues and pull requests created after discussing potential issues and features, which indicates that the workflow of discussion before working on issues or pull requests is used in some projects.
    Such a new workflow may result in new bug lifecycles that are different from those previously considered~\citep{10.1109/ICSE.2009.5070530}.
    However, we have also observed that some projects had stopped using GitHub Discussions. For example, the \textsf{dotnet/aspnetcore} project decided to use issues for discussions and disabled GitHub Discussions because they faced difficulties in labeling and assignment using it.\footnote{\url{https://github.com/dotnet/aspnetcore/issues/29935}}    
    What features are suitable for discussions may vary depending on the characteristics of projects, such as development histories, types of end users, maturity of the communities, etc. Since GitHub Discussions is still in beta, there is still room for improvement in terms of functionality. A better understanding of the required features and error/issue handling process depending on the characteristics of projects will be a challenge in the future.
    \item
    \textit{Further studies of human aspects.}
    As recent studies have shown, communication is an essential attribute of a great maintainer \citep{9402023}. A quantitative analysis of the relationship between the type and experience of developers and the discussion categories that require different skills such as technical, social, and managerial will be a future research challenge.
    \item
    \textit{Relationship between GitHub Discussions and other channels.}
    In addition to comparing to Stack Overflow, future work can also consider comparing GitHub Discussions to other communication channels used by open source projects, such as Gitter~\citep{Chatterjee2021, Sahar2021}, for example focusing on links between Gitter conversations and GitHub Discussions or comparing the channels in terms of their sentiment.
\end{itemize}

\section{Threats to Validity}
\label{sec:threats}

Threats to internal validity concern the possibility of introducing a threat due to a possible causal relationship between the treatment and the result of an analysis. As for sentiment and toxicity in GitHub Discussions and Stack Overflow posts, we performed the analysis at the platform level. However, we might have observed different distribution of sentiment and toxicity at the level of individual repositories, depending on the degree of awareness of the code of conduct of community members of a software project. 
Furthermore, we analysed and compared the sentiment polarity distribution in two dataset including 32,714 posts from GitHub Discussions and 99,918 posts from Stack Overflow, respectively. Specifically, we collected posts from Stack Overflow based on the presence of tags and owner names from GitHub projects in our dataset. The list of keywords (tags and name) was obtained via manual search on Stack Overflow. We acknowledge that manual search might introduce a threat to validity as well as replicability issues. To mitigate this threat, we include the list of keywords in our replication package.  
Furthermore, some projects did not have specific tags on Stack Overflow. Further replications could be conducted in the future on a larger set of data in order to include more projects, thus strengthening the representativeness of the sentiment analysis study.


Threats to construct validity concern the degree of accuracy to which the variables defined in a study measure the constructs of interest. As for analysis of sentiment, we operationalized developers' emotions 
by measuring the sentiment polarity of a text, that is its positive or negative orientation. However, emotion lexicon can convey a wide range of affective states or attitudes towards the interlocutor beyond polarity. 
As for participant types in Section~\ref{ssec:2.1}, although we identified three types, more fine-grained type identification including maintainers and collaborators could reveal project-specific patterns of discussion activities and discussion participation.


Threats to external validity concern the ability to generalize the findings in this study. As we study the usage of a beta feature, our data is derived from a short term (seven months).
Long-term studies may yield different conclusions. In addition, since we studied the use of GitHub Discussions in popular (based on the number of stargazers) and selected (by GitHub for beta testing) projects, our results could be biased to those specific communities. To address these issues, further research is needed based on various projects participating in GitHub Discussions after the feature is officially released.

\section{Related Work}
\label{sec:related}

\subsection{Communication during Software Development}

Discussions play a vital role in reaching a decision when doing code review. 
\cite{TsayFSE14} studied how developers used discussions to evaluate Pull Requests.
\cite{Hirao-tse-202003} studied the divergent nature during the code review, while \cite{EbertSANER2019} present a taxonomy of confusion that a reviewer may encounter when conducting a review. 
\cite{PascarellaCSCW18} present a taxonomy of information needs that help with conducting a review.
\cite{2020arXiv200909130S} studied the characteristics of the usage of a project-specific forum in the Eclipse project.
Other work has explored different aspects related to the analytics of review comments~\citep{HIRAO2019, RahmanMSR19}.
One aspect that could benefit is how review comments could be used for reviewer recommendation~\citep{JiangIST17}, where discussion recommendations can recommend appropriate team members to help answer generic questions from the community.
This can be beneficial for the community, especially for newcomers or even to help deepen the understanding between both users and developers of the software.

Besides code review, discussions also take place on online platforms such as Stack Overflow.
Previous work studied aspects such as who participates in such discussions~\citep{VasilescuCapiluppiOthers2012, MorrisonMurphyHill2015}, how frequently users contribute~\citep{WangLoDavidOthers2013}, how frequently discussed code snippets are copied between platforms~\citep{YangMartinsOthers2017, DBLP:journals/ese/BaltesD19}, and how content evolves on Q\&A websites~\citep{DBLP:conf/msr/BaltesDT008, Wang2018ReviseAnswers}.
From the study of discussions in issue tracking systems, Arya et al. reported various discussion categories in issue discussions, such as observed bug behaviour, bug reproduction, solution discussion, etc.~\citep{10.1109/ICSE.2019.00058}. We observed some categories overlap with categories found in GitHub Discussions, such as task progress, future plan, and social conversation. Since these discussions do not need to be closed with particular resolutions, GitHub Discussions could be an appropriate channel as mentioned by developers in Section~\ref{ssec:survey}.

From the point of a community knowledge base, source code comments have been found to document information specific to the associated code, such as technical debt~\citep{10.1109/ICSME.2014.31}, todo tasks~\citep{10.1145/1368088.1368123}, and algorithm~\citep{2019arXiv191006932I}. Similar to the analysis in Section~\ref{ssec:3.1}, referencing and connecting to external sources in source code comments had been analyzed~\citep{10.1109/ICSE.2019.00123}. A recent study identified specific cases of issues and technical debt, called on-hold self-admitted technical debt, which needs appropriate management of code considering the updates of issues~\citep{Maipradit2020,9252045}.
Although different knowledge is communicated and discussed in different communication channels, a new feature of GitHub Discussions could be a center of a community knowledge base connecting with other artefacts.


\subsection{Sentiment Polarity in Communication Channels}

Our sentiment analysis in Section~\ref{ssec:3.2} is motivated by the interest of the research community in sentiment analysis of developers' communication traces 
~\citep{IEEESoftwareEmotions2019,NOVIELLI2019JSS}. Specifically, researchers investigated  how developers share their emotions in the communication channels within collaborative development environments, including issue tracking systems (e.g., Jira)~\citep{Mantyla16, Ortu15}, software repository forges (e.g., GitHub)~\citep{Guzman14,Pletea14,Sinha16}, and technical Q\&A sites (e.g., Stack Overflow)~\citep{Uddin2017,CalefatoIST18,Lin2019}. What the above mentioned studies have in common is that they rely on sentiment analysis, -- i.e. the study of the subjectivity (neutral
vs. emotionally loaded) and polarity (positive vs. negative) of a
text~\citep{Pang07} -- to mine emotions and opinions from textual developer-generated content.
Beyond sentiment polarity, Gachechiladze et al. investigated the problem of identifying the target of anger expressions in collaborative software development~\citep{Gachechiladze17}. They distinguish between \textit{self}-directed emotions, such as sadness, which do not involve a negative attitude towards the interlocutor, hostile language expressing anger towards \textit{others}, and negative evaluation of \textit{objects}, such as tools or programming languages. 
More recently, Raman and colleagues proposed an approach to mine the toxicity of developers' online conversations towards identification and mitigation of unhealthy interactions in open source software development~\citep{Raman2020StressAB}. Toxicity has been also reported as a problem of Stack Overflow, which has been recognized as a hostile place especially for newcomers, women, non-native speakers, and others in marginalized groups by both users\footnote{\url{https://meta.stackoverflow.com/q/262791}} 
and community managers.\footnote{\url{https://stackoverflow.blog/2018/04/26/stack-overflow-isnt-very-welcoming-its-time-for-that-to-change/}} 

\section{Conclusion}
\label{sec:conclusion}

We conducted the first comprehensive mixed-methods study on how and why software developers adopt GitHub Discussions and how they perceive the new feature's usefulness.
We further studied how Discussions compare to existing channels such as GitHub Issues, Pull Requests, and Stack Overflow threads.
Based on an analysis of Discussions in 92 GitHub projects, a survey with developers already using the feature, and a comparison with related Stack Overflow posts, we derive the following recommendations
for project managers.
    (i) \textit{Set guidelines for participating in discussions,} as users sometimes find it difficult to choose an appropriate channel from Discussions, Issues, and Pull Requests. 
    (ii) \textit{Encourage core developers to participate in discussions,} as it is evident that the involvement of core developers is crucial in the successful use of GitHub Discussions. 
%

\bibliographystyle{spbasic}  
\bibliography{literature}

\begin{thebibliography}{68}
\providecommand{\natexlab}[1]{#1}
\providecommand{\url}[1]{{#1}}
\providecommand{\urlprefix}{URL }
\expandafter\ifx\csname urlstyle\endcsname\relax
  \providecommand{\doi}[1]{DOI~\discretionary{}{}{}#1}\else
  \providecommand{\doi}{DOI~\discretionary{}{}{}\begingroup
  \urlstyle{rm}\Url}\fi
\providecommand{\eprint}[2][]{\url{#2}}

\bibitem[{Abdellatif et~al(2020)Abdellatif, Badran, and
  Shihab}]{Abdellatif2020}
Abdellatif A, Badran K, Shihab E (2020) {MSRBot: Using bots to answer questions
  from software repositories}. Empir Softw Eng 25(3):1834--1863,
  \doi{10.1007/s10664-019-09788-5},
  \urlprefix\url{https://doi.org/10.1007/s10664-019-09788-5}

\bibitem[{Allamanis and Sutton(2013)}]{10.5555/2487085.2487098}
Allamanis M, Sutton C (2013) Why, when, and what: Analyzing stack overflow
  questions by topic, type, and code. In: Proc. of the 10th Working Conference
  on Mining Software Repositories, IEEE Press, MSR '13, pp 53--56

\bibitem[{Aranda and Venolia(2009)}]{10.1109/ICSE.2009.5070530}
Aranda J, Venolia G (2009) The secret life of bugs: Going past the errors and
  omissions in software repositories. In: Proc. of the 31st International
  Conference on Software Engineering, Association for Computing Machinery, New
  York, NY, USA, ICSE '09, p 298–308, \doi{10.1109/ICSE.2009.5070530},
  \urlprefix\url{https://doi.org/10.1109/ICSE.2009.5070530}

\bibitem[{Arya et~al(2019)Arya, Wang, Guo, and Cheng}]{10.1109/ICSE.2019.00058}
Arya D, Wang W, Guo JLC, Cheng J (2019) Analysis and detection of information
  types of open source software issue discussions. In: Proc. of the 41st
  International Conference on Software Engineering, IEEE Press, ICSE '19, pp
  454--464, \doi{10.1109/ICSE.2019.00058},
  \urlprefix\url{https://doi.org/10.1109/ICSE.2019.00058}

\bibitem[{Balali et~al(2020)Balali, Annamalai, Padala, Trinkenreich, Gerosa,
  Steinmacher, and Sarma}]{10.1145/3412569.3412571}
Balali S, Annamalai U, Padala HS, Trinkenreich B, Gerosa MA, Steinmacher I,
  Sarma A (2020) Recommending tasks to newcomers in oss projects: How do
  mentors handle it? In: Proc. of the 16th International Symposium on Open
  Collaboration, Association for Computing Machinery, New York, NY, USA,
  OpenSym '20, \doi{10.1145/3412569.3412571},
  \urlprefix\url{https://doi.org/10.1145/3412569.3412571}

\bibitem[{Baltes and Diehl(2019)}]{DBLP:journals/ese/BaltesD19}
Baltes S, Diehl S (2019) Usage and attribution of stack overflow code snippets
  in github projects. Empir Softw Eng 24(3):1259--1295,
  \doi{10.1007/s10664-018-9650-5},
  \urlprefix\url{https://doi.org/10.1007/s10664-018-9650-5}

\bibitem[{Baltes et~al(2018)Baltes, Dumani, Treude, and
  Diehl}]{DBLP:conf/msr/BaltesDT008}
Baltes S, Dumani L, Treude C, Diehl S (2018) Sotorrent: reconstructing and
  analyzing the evolution of stack overflow posts. In: Proc. of the 15th
  International Conference on Mining Software Repositories, {ACM}, MSR '18, pp
  319--330, \doi{10.1145/3196398.3196430},
  \urlprefix\url{https://doi.org/10.1145/3196398.3196430}

\bibitem[{Beyer and Pinzger(2016)}]{10.1145/2901739.2901750}
Beyer S, Pinzger M (2016) Grouping android tag synonyms on stack overflow. In:
  Proc. of the 13th International Conference on Mining Software Repositories,
  Association for Computing Machinery, New York, NY, USA, MSR '16, pp 430--440,
  \doi{10.1145/2901739.2901750},
  \urlprefix\url{https://doi.org/10.1145/2901739.2901750}

\bibitem[{Beyer et~al(2020)Beyer, Macho, {Di Penta}, and Pinzger}]{Beyer2020}
Beyer S, Macho C, {Di Penta} M, Pinzger M (2020) {What kind of questions do
  developers ask on Stack Overflow? A comparison of automated approaches to
  classify posts into question categories}. Empir Softw Eng 25(3):2258--2301,
  \doi{10.1007/s10664-019-09758-x}

\bibitem[{{Borges} et~al(2016){Borges}, {Hora}, and {Valente}}]{7816479}
{Borges} H, {Hora} A, {Valente} MT (2016) Understanding the factors that impact
  the popularity of github repositories. In: Proc. of the 32nd IEEE
  International Conference on Software Maintenance and Evolution, ICSME '16, pp
  334--344

\bibitem[{Braun and Clarke(2006)}]{braun2006using}
Braun V, Clarke V (2006) Using thematic analysis in psychology. Qualitative
  research in psychology 3(2):77--101

\bibitem[{Calefato et~al(2018{\natexlab{a}})Calefato, Lanubile, Maiorano, and
  Novielli}]{CalefatoEMSE18}
Calefato F, Lanubile F, Maiorano F, Novielli N (2018{\natexlab{a}}) Sentiment
  polarity detection for software development. Empir Softw Eng
  23(3):1352--1382, \doi{10.1007/s10664-017-9546-9},
  \urlprefix\url{https://doi.org/10.1007/s10664-017-9546-9}

\bibitem[{Calefato et~al(2018{\natexlab{b}})Calefato, Lanubile, and
  Novielli}]{CalefatoIST18}
Calefato F, Lanubile F, Novielli N (2018{\natexlab{b}}) How to ask for
  technical help? evidence-based guidelines for writing questions on stack
  overflow. Information {\&} Software Technology 94:186--207,
  \doi{10.1016/j.infsof.2017.10.009},
  \urlprefix\url{https://doi.org/10.1016/j.infsof.2017.10.009}

\bibitem[{Calefato et~al(2019)Calefato, Lanubile, Novielli, and
  Quaranta}]{EmtkSemotion2019}
Calefato F, Lanubile F, Novielli N, Quaranta L (2019) Emtk: The emotion mining
  toolkit. In: Proc. of the 4th International Workshop on Emotion Awareness in
  Software Engineering, IEEE Press, SEmotion ’19, pp 34--37,
  \doi{10.1109/SEmotion.2019.00014},
  \urlprefix\url{https://doi.org/10.1109/SEmotion.2019.00014}

\bibitem[{Chatterjee et~al(2021)Chatterjee, Damevski, and
  Pollock}]{Chatterjee2021}
Chatterjee P, Damevski K, Pollock L (2021) Automatic extraction of
  opinion-based q\&a from online developer chats. In: Proc. of the 43rd
  International Conference on Software Engineering (ICSE), IEEE, pp 1260--1272

\bibitem[{Cleary et~al(2013)Cleary, G{\'o}mez, Storey, Singer, and
  Treude}]{Cleary2013}
Cleary B, G{\'o}mez C, Storey MA, Singer L, Treude C (2013) Analyzing the
  friendliness of exchanges in an online software developer community. In:
  Proc. of the 6th International Workshop on Cooperative and Human Aspects of
  Software Engineering (CHASE), IEEE, pp 159--160

\bibitem[{Cleveland and Loader(1996)}]{10.1007/978-3-642-48425-4_2}
Cleveland WS, Loader C (1996) Smoothing by local regression: Principles and
  methods. In: H{\"a}rdle W, Schimek MG (eds) Statistical Theory and
  Computational Aspects of Smoothing, Physica-Verlag HD, Heidelberg, pp 10--49

\bibitem[{Cohen(1988)}]{cohen1988spa}
Cohen J (1988) {Statistical Power Analysis for the Behavioral Sciences}.
  Lawrence Erlbaum Associates

\bibitem[{Dias et~al(2021)Dias, Meirelles, Castor, Steinmacher, Wiese, and
  Pinto}]{9402023}
Dias E, Meirelles P, Castor F, Steinmacher I, Wiese I, Pinto G (2021) What
  makes a great maintainer of open source projects? In: Proc. of the 43rd
  International Conference on Software Engineering, ICSE '21, pp 982--994,
  \doi{10.1109/ICSE43902.2021.00093}

\bibitem[{{Ebert} et~al(2019){Ebert}, {Castor}, {Novielli}, and
  {Serebrenik}}]{EbertSANER2019}
{Ebert} F, {Castor} F, {Novielli} N, {Serebrenik} A (2019) Confusion in code
  reviews: Reasons, impacts, and coping strategies. In: Proc of the IEEE 26th
  International Conference on Software Analysis, Evolution and Reengineering,
  SANER '19, pp 49--60

\bibitem[{Fleiss and Cohen(1973)}]{Fleiss}
Fleiss JL, Cohen J (1973) The equivalence of weighted kappa and the intraclass
  correlation coefficient as measures of reliability. Educational and
  Psychological Measurement 33(3):613--619, \doi{10.1177/001316447303300309}

\bibitem[{Gachechiladze et~al(2017)Gachechiladze, Lanubile, Novielli, and
  Serebrenik}]{Gachechiladze17}
Gachechiladze D, Lanubile F, Novielli N, Serebrenik A (2017) Anger and its
  direction in collaborative software development. In: Proc. of the 39th
  International Conference on Software Engineering: New Ideas and Emerging
  Results Track, IEEE Press, ICSE-NIER ’17, pp 11--14,
  \doi{10.1109/ICSE-NIER.2017.18},
  \urlprefix\url{https://doi.org/10.1109/ICSE-NIER.2017.18}

\bibitem[{Giuffrida and Dittrich(2013)}]{giuffrida2013empirical}
Giuffrida R, Dittrich Y (2013) Empirical studies on the use of social software
  in global software development--a systematic mapping study. Information and
  Software Technology 55(7):1143--1164

\bibitem[{Guzman et~al(2014)Guzman, Az\'{o}car, and Li}]{Guzman14}
Guzman E, Az\'{o}car D, Li Y (2014) Sentiment analysis of commit comments in
  github: An empirical study. In: Proc. of the 11th Working Conf. on Mining
  Software Repositories, ACM, New York, NY, USA, MSR '14, pp 352--355,
  \doi{10.1145/2597073.2597118},
  \urlprefix\url{https://doi.org/10.1145/2597073.2597118}

\bibitem[{Guzzi et~al(2013)Guzzi, Bacchelli, Lanza, Pinzger, and
  Van~Deursen}]{guzzi2013communication}
Guzzi A, Bacchelli A, Lanza M, Pinzger M, Van~Deursen A (2013) Communication in
  open source software development mailing lists. In: Proc. of the 10th Working
  Conference on Mining Software Repositories, IEEE, MSR '13, pp 277--286

\bibitem[{Hata et~al(2015)Hata, Todo, Onoue, and
  Matsumoto}]{10.1109/CHASE.2015.9}
Hata H, Todo T, Onoue S, Matsumoto K (2015) Characteristics of sustainable oss
  projects: A theoretical and empirical study. In: Proc. of the IEEE/ACM 8th
  International Workshop on Cooperative and Human Aspects of Software
  Engineering, IEEE Computer Society, USA, CHASE '15, pp 15--21,
  \doi{10.1109/CHASE.2015.9},
  \urlprefix\url{https://doi.org/10.1109/CHASE.2015.9}

\bibitem[{Hata et~al(2019)Hata, Treude, Kula, and
  Ishio}]{10.1109/ICSE.2019.00123}
Hata H, Treude C, Kula RG, Ishio T (2019) 9.6 million links in source code
  comments: Purpose, evolution, and decay. In: Proc. of the 41st International
  Conference on Software Engineering, IEEE Press, ICSE '19, pp 1211--1221,
  \doi{10.1109/ICSE.2019.00123},
  \urlprefix\url{https://doi.org/10.1109/ICSE.2019.00123}

\bibitem[{Hata et~al(2021)Hata, Novielli, Baltes, Kula, and
  Treude}]{hideaki_hata_2021_4527166}
Hata H, Novielli N, Baltes S, Kula RG, Treude C (2021) {Research Artifact: An
  Exploratory Study of GitHub Discussions Early Adoption}.
  \doi{10.5281/zenodo.5026134},
  \urlprefix\url{https://doi.org/10.5281/zenodo.5026134}

\bibitem[{Hirao et~al(2019)Hirao, Kula, Ihara, and Matsumoto}]{HIRAO2019}
Hirao T, Kula RG, Ihara A, Matsumoto K (2019) Understanding developer
  commenting in code reviews. IEICE Transactions on Information and Systems
  E102.D(12):2423--2432

\bibitem[{Hirao et~al(2020)Hirao, McIntosh, Ihara, and
  Matsumoto}]{Hirao-tse-202003}
Hirao T, McIntosh S, Ihara A, Matsumoto K (2020) Code reviews with divergent
  review scores: An empirical study of the openstack and qt communities. IEEE
  Transactions on Software Engineering

\bibitem[{{Inokuchi} et~al(2019){Inokuchi}, {Sulistyo Nugroho},
  {Wattanakriengkrai}, {Konishi}, {Hata}, {Treude}, {Monden}, and
  {Matsumoto}}]{2019arXiv191006932I}
{Inokuchi} A, {Sulistyo Nugroho} Y, {Wattanakriengkrai} S, {Konishi} F, {Hata}
  H, {Treude} C, {Monden} A, {Matsumoto} K (2019) {From Academia to Software
  Development: Publication Citations in Source Code Comments}. arXiv e-prints
  arXiv:1910.06932, \eprint{1910.06932}

\bibitem[{Islam and Zibran(2017)}]{Islam17}
Islam MR, Zibran MF (2017) Leveraging automated sentiment analysis in software
  engineering. In: Proc. of the 14th International Conf. on Mining Software
  Repositories, IEEE Press, MSR ’17, pp 203--214, \doi{10.1109/MSR.2017.9},
  \urlprefix\url{https://doi.org/10.1109/MSR.2017.9}

\bibitem[{Jiang et~al(2017)Jiang, Yang, He, Blanc, and Zhang}]{JiangIST17}
Jiang J, Yang Y, He J, Blanc X, Zhang L (2017) Who should comment on this pull
  request? analyzing attributes for more accurate commenter recommendation in
  pull-based development. Inf Softw Technol pp 48--62

\bibitem[{Lin et~al(2019)Lin, Zampetti, Bavota, Di~Penta, and Lanza}]{Lin2019}
Lin B, Zampetti F, Bavota G, Di~Penta M, Lanza M (2019) Pattern-based mining of
  opinions in q\&a websites. In: Proc. of the 41st International Conference on
  Software Engineering, IEEE Press, ICSE ’19, pp 548--559,
  \doi{10.1109/ICSE.2019.00066},
  \urlprefix\url{https://doi.org/10.1109/ICSE.2019.00066}

\bibitem[{Maipradit et~al(2020{\natexlab{a}})Maipradit, Lin, Nagy, Bavota,
  Lanza, Hata, and Matsumoto}]{9252045}
Maipradit R, Lin B, Nagy C, Bavota G, Lanza M, Hata H, Matsumoto K
  (2020{\natexlab{a}}) Automated identification of on-hold self-admitted
  technical debt. In: Proc. of the IEEE 20th International Working Conference
  on Source Code Analysis and Manipulation, IEEE Computer Society, Los
  Alamitos, CA, USA, SCAM '20, pp 54--64, \doi{10.1109/SCAM51674.2020.00011},
  \urlprefix\url{https://doi.ieeecomputersociety.org/10.1109/SCAM51674.2020.00011}

\bibitem[{Maipradit et~al(2020{\natexlab{b}})Maipradit, Treude, Hata, and
  Matsumoto}]{Maipradit2020}
Maipradit R, Treude C, Hata H, Matsumoto K (2020{\natexlab{b}}) {Wait for it:
  identifying “On-Hold” self-admitted technical debt}. Empir Softw Eng
  25(5):3770--3798, \doi{10.1007/s10664-020-09854-3},
  \urlprefix\url{https://doi.org/10.1007/s10664-020-09854-3}

\bibitem[{M\"{a}ntyl\"{a} et~al(2016)M\"{a}ntyl\"{a}, Adams, Destefanis,
  Graziotin, and Ortu}]{Mantyla16}
M\"{a}ntyl\"{a} M, Adams B, Destefanis G, Graziotin D, Ortu M (2016) Mining
  valence, arousal, and dominance: Possibilities for detecting burnout and
  productivity? In: Proc. of the 13th International Conf. on Mining Software
  Repositories, ACM, New York, NY, USA, MSR ’16, pp 247--258,
  \doi{10.1145/2901739.2901752},
  \urlprefix\url{https://doi.org/10.1145/2901739.2901752}

\bibitem[{Morrison and Murphy-Hill(2015)}]{MorrisonMurphyHill2015}
Morrison P, Murphy-Hill E (2015) {Is programming knowledge related to age? An
  exploration of Stack Overflow}. In: {Di Penta} M, Pinzger M, Robbes R (eds)
  {12th Working Conference on Mining Software Repositories (MSR 2015)}, {IEEE
  Computer Society}, Florence, Italy, pp 69--72

\bibitem[{Munaiah et~al(2017)Munaiah, Kroh, Cabrey, and
  Nagappan}]{DBLP:journals/ese/MunaiahKCN17}
Munaiah N, Kroh S, Cabrey C, Nagappan M (2017) Curating github for engineered
  software projects. Empir Softw Eng 22(6):3219--3253,
  \doi{10.1007/s10664-017-9512-6},
  \urlprefix\url{https://doi.org/10.1007/s10664-017-9512-6}

\bibitem[{Murgia et~al(2014)Murgia, Tourani, Adams, and Ortu}]{Murgia2014}
Murgia A, Tourani P, Adams B, Ortu M (2014) Do developers feel emotions? an
  exploratory analysis of emotions in software artifacts. In: Proc. of the 11th
  Working Conf. on Mining Software Repositories, ACM, New York, NY, USA, MSR
  '14, pp 262--271, \doi{10.1145/2597073.2597086},
  \urlprefix\url{https://doi.org/10.1145/2597073.2597086}

\bibitem[{{Novielli} and {Serebrenik}(2019)}]{IEEESoftwareEmotions2019}
{Novielli} N, {Serebrenik} A (2019) Sentiment and emotion in software
  engineering. IEEE Software 36(5):6--23, \doi{10.1109/MS.2019.2924013}

\bibitem[{Novielli et~al(2018)Novielli, Girardi, and
  Lanubile}]{10.1145/3196398.3196403}
Novielli N, Girardi D, Lanubile F (2018) A benchmark study on sentiment
  analysis for software engineering research. In: Proc. of the 15th
  International Conference on Mining Software Repositories, Association for
  Computing Machinery, New York, NY, USA, MSR '18, pp 364--375,
  \doi{10.1145/3196398.3196403},
  \urlprefix\url{https://doi.org/10.1145/3196398.3196403}

\bibitem[{Novielli et~al(2019)Novielli, Begel, and Maalej}]{NOVIELLI2019JSS}
Novielli N, Begel A, Maalej W (2019) Introduction to the special issue on
  affect awareness in software engineering. Journal of Systems and Software
  148:180--182, \doi{https://doi.org/10.1016/j.jss.2018.11.016},
  \urlprefix\url{http://www.sciencedirect.com/science/article/pii/S0164121218302504}

\bibitem[{Ortu et~al(2015)Ortu, Adams, Destefanis, Tourani, Marchesi, and
  Tonelli}]{Ortu15}
Ortu M, Adams B, Destefanis G, Tourani P, Marchesi M, Tonelli R (2015) Are
  bullies more productive? empirical study of affectiveness vs. issue fixing
  time. In: Proc. of the 12th Working Conf. on Mining Software Repositories,
  IEEE Press, MSR ’15, pp 303--313

\bibitem[{Pang and Lee(2008)}]{Pang07}
Pang B, Lee L (2008) Opinion mining and sentiment analysis. Foundations and
  Trends in Information Retrieval 2(1-2):1--135, \doi{10.1561/1500000011},
  \urlprefix\url{https://doi.org/10.1561/1500000011}

\bibitem[{Pascarella et~al(2018)Pascarella, Spadini, Palomba, Bruntink, and
  Bacchelli}]{PascarellaCSCW18}
Pascarella L, Spadini D, Palomba F, Bruntink M, Bacchelli A (2018) {Information
  Needs in Contemporary Code Review}. In: Proc. of the 21st ACM Conference on
  Computer Supported Cooperative Work, CSCW '18, vol~2, pp 135:1--135:27

\bibitem[{Pletea et~al(2014)Pletea, Vasilescu, and Serebrenik}]{Pletea14}
Pletea D, Vasilescu B, Serebrenik A (2014) Security and emotion: Sentiment
  analysis of security discussions on github. In: Proc. of the 11th Working
  Conf. on Mining Software Repositories, ACM, New York, NY, USA, MSR '14, pp
  348--351, \doi{10.1145/2597073.2597117},
  \urlprefix\url{https://doi.org/10.1145/2597073.2597117}

\bibitem[{Potdar and Shihab(2014)}]{10.1109/ICSME.2014.31}
Potdar A, Shihab E (2014) An exploratory study on self-admitted technical debt.
  In: Proc. of the 2014 IEEE International Conference on Software Maintenance
  and Evolution, IEEE Computer Society, USA, ICSME '14, pp 91--100,
  \doi{10.1109/ICSME.2014.31},
  \urlprefix\url{https://doi.org/10.1109/ICSME.2014.31}

\bibitem[{Rahman et~al(2017)Rahman, Roy, and Kula}]{RahmanMSR19}
Rahman MM, Roy CK, Kula RG (2017) Predicting usefulness of code review comments
  using textual features and developer experience. In: Proc. of the 14th
  International Conference on Mining Software Repositories, MSR '17, pp
  215--226

\bibitem[{Raman et~al(2020)Raman, Cao, Tsvetkov, K\"{a}stner, and
  Vasilescu}]{Raman2020StressAB}
Raman N, Cao M, Tsvetkov Y, K\"{a}stner C, Vasilescu B (2020) Stress and
  burnout in open source: Toward finding, understanding, and mitigating
  unhealthy interactions. In: Proc. of the ACM/IEEE 42nd International
  Conference on Software Engineering: New Ideas and Emerging Results,
  Association for Computing Machinery, New York, NY, USA, ICSE-NIER '20, pp
  57--60, \doi{10.1145/3377816.3381732},
  \urlprefix\url{https://doi.org/10.1145/3377816.3381732}

\bibitem[{Robillard and Treude(2020)}]{Robillard2020}
Robillard MP, Treude C (2020) Understanding wikipedia as a resource for
  opportunistic learning of computing concepts. In: Proc. of the 51st ACM
  Technical Symposium on Computer Science Education, Association for Computing
  Machinery, New York, NY, USA, SIGCSE '20, pp 72–--78,
  \doi{10.1145/3328778.3366832},
  \urlprefix\url{https://doi.org/10.1145/3328778.3366832}

\bibitem[{Rosen and Shihab(2016)}]{10.1007/s10664-015-9379-3}
Rosen C, Shihab E (2016) What are mobile developers asking about? a large scale
  study using stack overflow. Empir Softw Eng 21(3):1192--1223,
  \doi{10.1007/s10664-015-9379-3},
  \urlprefix\url{https://doi.org/10.1007/s10664-015-9379-3}

\bibitem[{Sahar et~al(2021)Sahar, Hindle, and Bezemer}]{Sahar2021}
Sahar H, Hindle A, Bezemer CP (2021) How are issue reports discussed in gitter
  chat rooms? Journal of Systems and Software 172:110852

\bibitem[{Sinha et~al(2016)Sinha, Lazar, and Sharif}]{Sinha16}
Sinha V, Lazar A, Sharif B (2016) Analyzing developer sentiment in commit logs.
  In: Proc. of the 13th International Conf. on Mining Software Repositories,
  ACM, New York, NY, USA, MSR ’16, pp 520--523,
  \doi{10.1145/2901739.2903501},
  \urlprefix\url{https://doi.org/10.1145/2901739.2903501}

\bibitem[{Steinmacher et~al(2014)Steinmacher, Graciotto~Silva, Gerosa, and
  Redmiles}]{Steinmacher:2014}
Steinmacher I, Graciotto~Silva MA, Gerosa MA, Redmiles D (2014) A systematic
  literature review on the barriers faced by newcomers to open source software
  projects. Information and Software Technology 59:67--85,
  \doi{10.1016/j.infsof.2014.11.001}

\bibitem[{Steinmacher et~al(2019)Steinmacher, Treude, and Gerosa}]{8254320}
Steinmacher I, Treude C, Gerosa MA (2019) Let me in: Guidelines for the
  successful onboarding of newcomers to open source projects. IEEE Software
  36(4):41--49, \doi{10.1109/MS.2018.110162131}

\bibitem[{Storey et~al(2008)Storey, Ryall, Bull, Myers, and
  Singer}]{10.1145/1368088.1368123}
Storey MA, Ryall J, Bull RI, Myers D, Singer J (2008) Todo or to bug: Exploring
  how task annotations play a role in the work practices of software
  developers. In: Proc. of the 30th International Conference on Software
  Engineering, Association for Computing Machinery, New York, NY, USA, ICSE
  '08, pp 251--260, \doi{10.1145/1368088.1368123},
  \urlprefix\url{https://doi.org/10.1145/1368088.1368123}

\bibitem[{Storey et~al(2016)Storey, Zagalsky, Figueira~Filho, Singer, and
  German}]{storey2016social}
Storey MA, Zagalsky A, Figueira~Filho F, Singer L, German DM (2016) How social
  and communication channels shape and challenge a participatory culture in
  software development. IEEE Transactions on Software Engineering
  43(2):185--204

\bibitem[{{Sulistyo Nugroho} et~al(2020){Sulistyo Nugroho}, {Islam}, {Nakasai},
  {Rehman}, {Hata}, {Gaikovina Kula}, {Nagappan}, and
  {Matsumoto}}]{2020arXiv200909130S}
{Sulistyo Nugroho} Y, {Islam} S, {Nakasai} K, {Rehman} I, {Hata} H, {Gaikovina
  Kula} R, {Nagappan} M, {Matsumoto} K (2020) {Sustaining a Healthy Ecosystem:
  Participation, Discussion, and Interaction in Eclipse Forums}. arXiv e-prints
  arXiv:2009.09130, \eprint{2009.09130}

\bibitem[{Thelwall et~al(2010)Thelwall, Buckley, Paltoglou, Cai, and
  Kappas}]{Thelwall2010}
Thelwall M, Buckley K, Paltoglou G, Cai D, Kappas A (2010) Sentiment strength
  detection in short informal text. J Am Soc Inf Sci Technol 61(12):2544--2558

\bibitem[{Treude et~al(2011)Treude, Barzilay, and
  Storey}]{10.1145/1985793.1985907}
Treude C, Barzilay O, Storey MA (2011) How do programmers ask and answer
  questions on the web? (nier track). In: Proc. of the 33rd International
  Conference on Software Engineering, Association for Computing Machinery, New
  York, NY, USA, ICSE '11, pp 804--807, \doi{10.1145/1985793.1985907},
  \urlprefix\url{https://doi.org/10.1145/1985793.1985907}

\bibitem[{Tsay et~al(2014)Tsay, Dabbish, and Herbsleb}]{TsayFSE14}
Tsay J, Dabbish L, Herbsleb J (2014) Let's talk about it: Evaluating
  contributions through discussion in github. In: Proc. of the 22nd ACM SIGSOFT
  International Symposium on Foundations of Software Engineering, FSE '14, pp
  144--154

\bibitem[{Uddin and Khomh(2017)}]{Uddin2017}
Uddin G, Khomh F (2017) Opiner: An opinion search and summarization engine for
  apis. In: Proc. of the 32nd IEEE/ACM International Conf. on Automated
  Software Engineering, IEEE Press, ASE '17, pp 978--983

\bibitem[{Vasilescu et~al(2012)Vasilescu, Capiluppi, and
  Serebrenik}]{VasilescuCapiluppiOthers2012}
Vasilescu B, Capiluppi A, Serebrenik A (2012) {Gender, Representation and
  Online Participation: A Quantitative Study of StackOverflow}. In: Aberer K,
  Flache A, Jager W, Liu L, Tang J, Gueret C (eds) Proc. of the 4th
  International Conference on Social Informatics, Springer, Lausanne,
  Switzerland, SocInfo '12, pp 332--338

\bibitem[{Viera and Garrett(2005)}]{Viera2005}
Viera AJ, Garrett JM (2005) Understanding interobserver agreement: the kappa
  statistic. Family medicine 37 5:360--3

\bibitem[{Wang et~al(2013)Wang, {Lo David}, and Jiang}]{WangLoDavidOthers2013}
Wang S, {Lo David}, Jiang L (2013) {An empirical study on developer
  interactions in StackOverflow}. In: Shin SY, Maldonado JC (eds) Proc. of the
  28th Annual ACM Symposium on Applied Computing, ACM, Coimbra, Portugal, SAC
  '13, pp 1019--1024

\bibitem[{{Wang} et~al(2018){Wang}, {Chen}, and
  {Hassan}}]{Wang2018ReviseAnswers}
{Wang} S, {Chen} TP, {Hassan} AE (2018) How do users revise answers on
  technical q\&a websites? a case study on stack overflow. IEEE Transactions on
  Software Engineering 46(9):1024--1038

\bibitem[{Yang et~al(2017)Yang, Martins, Saini, and
  Lopes}]{YangMartinsOthers2017}
Yang D, Martins P, Saini V, Lopes CV (2017) {Stack Overflow in Github: Any
  Snippets There?} In: Gonzalez-Barahona JM, Hindle A, Tan L (eds) Proc. of the
  14th International Conference on Mining Software Repositories, {IEEE Computer
  Society}, Buenos Aires, Argentina, MSR '17, pp 280--290

\end{thebibliography}
\end{sloppy}
\end{document}